\documentstyle[10pt,aaspp4,psfig]{article}
\tightenlines
\font\cap=cmcsc10
\def\ni{\noindent}        

\def\hi{\noindent \hangindent=2.5em}
\def\hour{^{\rm h}}

\def\degree{{\rm\,degree}}
\def\pixel{{\rm\,pixel}}

\def\kms{{\rm\,km/s}}
\def\kpc{{\rm\,kpc}}
\def\Mpc{{\rm\,Mpc}}
\def\mpc{{\rm\,Mpc}}

\def\msun{{\rm\,M_\odot}}
\def\lsun{{\rm\,L_\odot}}

\def\surfb{{\rm\,mag/arcsec^2}}

\def\aj{{\it A.~J.}, }  
\def\apj{{\it Ap.~J.}, }  
\def\apjs{{\it Ap.~J.~Suppl.}, }  
\def\mn{{\it M.N.R.A.S.}, }      
\def\nat{{\it Nature}, }      
\def\aa{{\it Astr.~Ap.}, }     
\def\aasup{{\it Astr.~Ap.~Suppl.}, }     

\def\spose#1{\hbox to 0pt{#1\hss}}
\def\lta{\mathrel{\spose{\lower 3pt\hbox{$\mathchar"218$}}
     \raise 2.0pt\hbox{$\mathchar"13C$}}}
\def\gta{\mathrel{\spose{\lower 3pt\hbox{$\mathchar"218$}}
     \raise 2.0pt\hbox{$\mathchar"13E$}}}

\def\clock{\count0=\time \divide\count0 by 60
     \count1=\count0 \multiply\count1 by -60 \advance\count1 by \time
     \number\count0:\ifnum\count1<10{0\number\count1}\else\number\count1\fi}

\begin{document}

\title{The Number Density of Low Surface Brightness Galaxies \\
with $23<\mu_0<25\,V\surfb$}

\author{Julianne J. Dalcanton\altaffilmark{1,2,3}}
\affil{Observatories of the Carnegie Institution of Washington,
       813 Santa Barbara St, Pasadena, CA 91101 \\
		\& \\
	Princeton University Observatory, Princeton, NJ 08544}
\author{David N. Spergel, James E. Gunn\altaffilmark{4}}
\affil{Princeton University Observatory,
       Princeton, NJ 08544}
\author{Maarten Schmidt}
\affil{Palomar Observatory,  California Institute of Technology, Pasadena, CA 91125}
\author{Donald P. Schneider\altaffilmark{5}}
\affil{Department of Astronomy \& Astrophysics, The Pennsylvania State
University, University Park, PA 16802.}
\altaffiltext{1}{e-mail address: jd@ociw.edu}
\altaffiltext{2}{Hubble Fellow}
\altaffiltext{3}{Visiting Astronomer, Kitt Peak National Observatory,
National Optical Astronomy Observatories, operated by the Association of
Universities for Research in Astronomy, Inc., under contract with the National
Science Foundation.}
\altaffiltext{4}{Visiting Associate, Palomar Observatory}
\altaffiltext{5}{Guest Investigator, Palomar Observatory}

\begin{abstract}

We present results of a large area CCD survey for low surface
brightness galaxies (LSBs) that reaches central surface brightnesses
of $25\surfb$ in $V$.  We have analyzed $17.5\,\degree^2$ of transit
scan data, and identified a statistical subset of 7 pure disk LSB's
with central surface brightnesses fainter than $\mu_0=23\,V\surfb$ and
with angular exponential scale lengths larger than
$\alpha=2.5\arcsec$.  The LSB detection is entirely automated, and the
selection efficiency of the survey is well-quantified.  After
correcting for the selection efficiency, we find a surface density of
$4.1^{+2.6}_{-2.1}\, {\rm galaxies / degree^2}$ for LSBs in the
considered range of $\mu_0$ and $\alpha$ (90\% confidence levels),
with the largest correction being due to the area lost behind bright
stars, and the difficulty in detecting LSBs with small angular sizes.

We have measured redshifts to the final sample of LSBs, and find them
to be at distances comparable to those probed by large galaxy
catalogs, and to have intrinsic scale lengths of $1.7-3.6\,{\rm
h}_{50}^{-1}\kpc$, also comparable to normal galaxies.  We use the
redshifts and the selection efficiency to calculate the number density
in LSBs with $23<\mu_0<25\,V\surfb$ and find ${\cal N} =
0.01^{+0.006}_{-0.005} \, {\rm galaxies \,h_{50}^3 Mpc^{-3}}$, with
90\% confidence.  The measurement of the absolute number density of
LSBs probably represents a lower limit, due to very strong biases
against LSBs with bulges or edge-on LSBs in our sample.  Comparing the
LSB number density to the number density of normal galaxies with
either similar scale lengths or similar luminosities, we find that the
number density of LSBs with $23<\mu_0<25\,V\surfb$ is comparable to or
greater than the number density of normal galaxies.  The luminosity
density in LSBs is comparable to the luminosity density of normal
galaxies with similar luminosities, but is a factor of 3-10 smaller
than the luminosity density of normal galaxies with similar scale
lengths.  The relative LSB number density and luminosity density agree
well with the theoretical predictions of Dalcanton et al.\ (1997).
The redshift-space distribution of the LSBs suggests that the trend
for low surface brightness galaxies to have weak small-scale
correlations may continue to the fainter surface brightnesses
covered in this survey.

\end{abstract}

\section{Introduction}

During the past twenty years, there has been a developing appreciation
of the strong biases against finding galaxies of low surface
brightness.  These biases arise because the night sky
is not particularly dark.  Airglow, zodiacal light, and
undetected stars and galaxies combine to create a optical background
whose surface brightness is as bright as the extrapolated central
surface brightnesses of large spiral galaxy disks.  With such a bright
background, the ability to detect a galaxy depends not only upon the
integrated luminosity of the galaxy, but also upon the contrast with
which the galaxy stands out above the Poisson fluctuations in the
background.  A compact, high-surface brightness galaxy might be quite
easy to detect, while another galaxy with the same total luminosity
but with a much more extended, low-surface brightness structure would
be difficult to find.

While astronomers routinely consider the limiting magnitude of their
galaxy catalogs, only recently have they begun to consider the ways in
which surface brightness selection effects shape existing galaxy
catalogs.  As an example, Freeman (1970) had initially showed that
spiral galaxies share the same central surface brightness, with very
little dispersion: $\mu_0 = 21.7 \pm 0.3\,B\surfb$.  Disney (1976),
however, convincingly argued that because of the limiting surface
brightness of the sample, one would not expect to have detected galaxies with
lower surface brightnesses, and thus that there was no
particularly compelling reason to believe that the Freeman law
reflected the intrinsic properties of spiral galaxies.
While Allen \& Shu (1979) later showed that the cutoff at bright surface
brightnesses does indeed reflect an actual fall-off
in the number of galaxies with increasing surface brightness, they
concurred with Disney's appraisal of the possible role of selection biases in
shaping the faint end of Freeman's surface brightness distribution.
Thus, Disney demonstrated that what was thought to be a general
property of disk galaxies could in large part be explained by
selection biases against finding low surface brightness galaxies
(LSBs).

Disney's paper was the genesis of the modern efforts both toward
understanding the role that surface brightness selection effects play
in shaping existing galaxy catalogs and toward overcoming those biases
in newer surveys.  Although Disney's work strongly suggested that
selection biases could reproduce the Freeman law, it presupposes that
there exists a population of low surface brightness galaxies to be
biased against.  It may have been that there really were almost no
galaxies with central surface brightnesses fainter than $22\surfb$.
Over the following years and extending through the present, however,
there has been an enormous body of observational work that has
conclusively demonstrated the existence of low surface brightness
galaxies.  Indeed, every time a new survey has been extend to fainter
surface brightnesses, new LSBs have been discovered.

Previous surveys which have been sensitive to LSBs have fallen into
two basic types: large area photographic surveys and deep, small area
CCD surveys.  Of the first type, the older, diameter-limited Upsalla
General Catalog of Galaxies (UGC) (Nilson 1973) had been the workhorse
through most of the 70's and 80's.  It has been recently supplemented
by a catalog of LSBs selected visually from the Second Palomar Sky
Survey using the same selection criteria as the UGC catalog but with
deeper plates (Schombert et al.\ 1992, Schombert \& Bothun 1988,
herein referred to as the ``POSS-II'' catalog), and now by the APM LSB
survey which selected LSBs automatically and by eye using scanned
plates from the APM galaxy survey (Impey et al.\ 1996).  Other deep
photographic work has been used to search for LSBs in clusters (Impey
et al.\ 1988, Davies et al.\ 1988, Irwin et al.\ 1990, Ferguson \&
Sandage 1988, Sandage \& Binggeli 1984).  With the exception of the
deeper cluster surveys, the photographic work finds very few galaxies
whose peak surface brightness is fainter than $\mu=24\surfb$ in $B$
(corresponding to $\approx23.5\surfb$ in $V$, using mean colors from
de Blok 1995, McGaugh 1994), unless they also have high surface
brightness bulges.

The second type of LSB survey, the CCD surveys, have concentrated
primarily on identifying LSBs within clusters of galaxies (Turner et
al.\ 1993).  Due to their small area, these surveys are restricted to
finding galaxies of small angular size, which leads to a large degree
of confusion between intrinsically low surface brightness galaxies and
either high-redshift normal galaxies which appear to have low surface
brightness because of $(1+z)^{-4}$ cosmological dimming, or pairs of
barely resolved galaxies.  A similar field survey by Schwartzenberg et
al (1995) also suffers from the same problems, although to a slightly
lesser extent due to a larger angular size cutoff.  However, the
difficulties in untangling the true surface brightness and angular
size distributions are compounded by effects of poor seeing and the
large pixel size of these surveys.  There has also been a promising
recent foray into using CCDs to survey for faint field LSBs through
cross-correlating an image with a model galaxy profile (Davies et al.\
1994).  However, the areas surveyed so far are still too small to do
more than to verify the promise of this method.

In spite of the limitations, existing surveys have
begun to advance our understanding of the density of the LSB
population.  They have both proven the existence of LSBs, and provided
a rich data set for studying the properties of individual LSBs.  They
have also begun to yield quantitative results about the size of the
LSB population.  First, several existing photographic and CCD cluster
surveys have be used to estimate the relative number density as a
function of surface brightness.  McGaugh (1996) has used the selection
criteria of the Davies (1990) sample to estimate the relative survey
volume as a function of central surface brightness, and shown that the
distribution of disk surface brightnesses is very broad.  The derived
distribution of surface brightnesses cuts off sharply brighter than
the Freeman surface brightness value, but falls only slowly with
decreasing central surface brightness; by $\mu_0(B)=23\surfb$, the
number density drops by only a factor of 2-4, but is statistically
consistent with a flat distribution in surface brightness.  However,
this analysis assumes that the distribution of intrinsic disk scale
lengths is independent of surface brightness, which does not seem to
be true (i.e.\ the increase in the maximum scale length with
decreasing surface brightness -- see de Jong \& van der Kruit 1994,
for example).  Thus, there may be systematic offsets in the accessible
survey volume as a function of surface brightness, leading to
substantial uncertainties in the relative number density as a function
of surface brightness.

More recently, the large APM LSB survey has reconstructed the
luminosity function of LSBs with central surface brightnesses of
$22\surfb<\mu_0(B)\lta24\surfb$, using a detailed quantification of the
automated detection algorithm (Sprayberry 1997).  The survey has
332 galaxies with redshifts, and has $\sim20$\% completeness in
redshift identification for galaxies with $\mu_0(B) > 23\surfb$.
They find that, when compared to the CfA survey luminosity function
(Marzke et al 1994) which covers a nearly disjoint surface brightness
range, the LSBs in their survey add rougly a third to the known luminosity
density in galaxies with $M_B < -13$, and more than double the known
luminosity density contributed by known galaxies with Irr/Im morphologies.

In the survey described in this paper, we attempt to combining the
best features of previous surveys, and to push observations to lower
surface brightnesses than have been previously surveyed in the field.
We use CCDs to observe a large area of the sky, obtain extremely
accurate flat-fielding through the use of drift scanning, use
completely automated detection to allow for a complete and detailed
understanding of the selection biases in the survey, and search for
galaxies with large angular sizes to reduce the background confusion
and to identify LSBs which are most comparable to well-cataloged
normal galaxies in large local surveys.  In the following section,
\S\ref{strategy}, we outline the basic strategy of our survey, and
discuss how it is optimized for reaching the above goals.  In
\S\ref{datareduction} we describe the data we have used, and the data
processing and selection criteria in detail.  In \S\ref{properties} we
discuss the properties of the entire sample of objects detected in the
survey, and in \S\ref{lsbsamp} we focus our attention on the
properties of the subset of LSB galaxies.  Finally, in
\S\ref{numberden} we use the LSB subsample to measure the surface
density, number density, and luminosity density of LSB galaxies, and
compare to the population to normal galaxies.  We discuss our
estimation of errors in Appendix A.  In Appendix B is a general
discussion of reconstructing the number and luminosity density of
angular diameter limited samples from the observed surface density.
We use $H_0=50\kms/\Mpc$ throughout.

\section{Strategy} \label{strategy}

There are two essential elements to the survey which we have carried
out, namely the use of time-delay-and-integrate (TDI) observations to
achieve extremely accurate flat-fielding over large areas and the use
of optimal smoothing filters to enhance the detection of large
LSBs. In a TDI scan (also known as ``strip'',``drift'', or ``transit''
scanning), the telescope is held fixed, while the sky drifts across
the field of view of the CCD camera, while lines of charge are stepped
from row to row of the CCD to keep track with the sky.  Drift scanning
averages out the flat-fielding variations along the columns of the
chip, and means that each line of the resulting strip of data has the
same flat-fielding (or ``flat-lining'') characteristics, up to the
temporal stability of the chip response.  Millions of lines of the
data can be combined to create an extremely accurate one-dimensional
sky flat for the entire data set, much in the same manner that
shift-and-stare techniques can be used to make two-dimensional sky
flats.  Drift-scanning therefore provides an extremely efficient way
of using CCDs to observe large areas of the sky while providing
excellent flat fielding.

With the advantage of extremely accurate large scale flat fielding, we
can smooth the data over large scales to reveal low surface brightness
galaxies which are typically at the limits of detectability, due to
Poisson pixel-to-pixel fluctuations in the sky level.  To increase
sensitivity to galaxies whose surface brightnesses are comparable to
the sky brightness, the pixel-to-pixel noise can also be reduced by
averaging together many pixels, increasing the effective area of the
pixels.  While this decrease in noise may be achieved by observing with
large CCD pixels or by binning down the charge on the chip, it may
also be reached by smoothing high resolution data.

There are several advantages to using smoothing to reduce the
pixel-to-pixel noise.  First, the smoothing may be done with a filter
that is matched to the shape and size of the galaxy for which one is
looking; this produces the maximum gain in signal-to-noise for the
targeted galaxies.  Second, because the original, high-resolution data
still exists after smoothing, no information is really lost; the
original data can provide additional information about the sources of any
low-surface brightness features detected after smoothing.  If the
smoothing were to have been done by binning the CCD during readout,
and not during post-processing, the increase in signal-to-noise would
have been optimized for square galaxies, not realistic ones, and most
importantly, there would be much more confusion about the source of
any fluctuations.  Finally, because smoothing allows any filter to be
used, one data set may potentially be used to independently survey for
galaxies of many different shapes and angular sizes.  To avoid
contributions from bright foreground objects, we have cleaned images
of bright foreground objects, such that subsequent smoothing truly
recovers large scale fluctuations in the optical background.

	Because the known population of LSBs seems to be well fit by
exponential disks (de Jong 1996, de Blok et al 1995, McGaugh \& Bothun
1994, Knezek 1993), we have chosen to use an exponential profile as
the filter kernel.  We have smoothed with an exponential filter of
scale length $5\arcsec$, which drops the level of fluctuations by over
a factor of ten, a gain that would only have been achieved with over a
hundred-fold increase in exposure time.  By choosing such a large
scale length, we eliminate any confusion between low surface
brightness galaxies and distant, cosmologically-dimmed high surface
brightness galaxies.  For example, galaxies with an apparent scale length of
$\approx5\arcsec$ are likely to be within $200\mpc$ ($z=0.03$,
assuming an intrinsic scale length of $5\,h_{50}\kpc$) and are near enough
that $(1+z)^4$ surface brightness dimming is negligible.  This choice
of scale length also identifies LSBs within a volume comparable to
that which is occupied by NGC and UGC galaxies; a normal galaxy with
a $5\arcsec$ exponential scale length has an angular diameter of
close to $1\arcmin$ (assuming $\mu_0=21.5\surfb$ and $\mu_{lim}=26.5$),
similar to the angular diameter limit of large photographic surveys.

To identify candidate LSBs, we identify all large regions of connected
pixels in the smoothed image which are significantly above the level of
the remaining fluctuations in the sky.  The entire process --
cleaning, smoothing, and searching -- is entirely automated, and is
repeated with artificial galaxies added to the data to allow a
thorough investigation of the selection biases of the survey.  We then
visually inspect all candidates to identify the subset which are LSBs.
We perform additional artificial galaxy tests to test the
limits of our ability to classify LSBs.  We use these limits to
further restrict the LSB subset to the region where our detection and
classification efficiencies are well-determined.  Finally, we
measure redshifts for the final LSB subset, allowing us to measure
the absolute number density of the LSBs.

\section{Data Processing}				\label{datareduction}

\subsection{Imaging Data}				\label{data}

	The survey utilizes existing imaging data, originally taken
for the Palomar Transit Grism Survey (Schneider, Schmidt, \& Gunn
1994; hereafter SSG) for high redshift quasars.  We will give a brief
summary of the initial data processing of this data set; see SSG for a
detailed description of the survey.  The data were taken with the
4-Shooter camera (Gunn et al.\ 1987) at the Cassegrain focus of the
Palomar 200-inch telescope.  The 4-Shooter detector consists of 4
800x800 Texas Instrument CCDs, arranged in a 2x2 grid, with a pixel
scale of $0.335 \arcsec / \pixel$.  The telescope was operated in
transit mode, holding the telescope fixed and allowing the sky to
drift across the CCDs while the CCD was clocked to keep pace with the
passage of the sky.  The drift scan produces four very long images at
a fixed declination covering a wide range in right ascension, each
strip corresponding to one of the four chips.  The strip is cut into
400x1600 images with 120 rows of overlap between adjacent images.
Because of difficulties with the high data flow rate, the CCDs were
double-clocked and the columns rebinned by a factor of 2, to increase
the effect of pixel scale to $0.67 \arcsec / \pixel $.  The seeing was
$1.5-3 \arcsec$.  The two leading CCDs used the $F555W$ ``wide $V$''
filter, while the two trailing CCDs used $F785LP$ (Griffiths 1990).
Following Postman et al.\ (1996), we will use the shorthand notation
of $V_4$ and $I_4$ for the combination of the HST filters and the
4-shooter response function.  Only the $V_4$ strips were used, because
of the greater susceptibility of the $I_4$ band to fringing and to low
level fluctuations in the sky brightness.

	We specifically use the QR and MN strips from the SSG survey,
observed in April 1987 and 1988, respectively.
(The two-letter notation reflects the fact that each strip is actually
composed of two separate strips, corresponding to the two halves of
the 4-shooter's CCD array.  We analyze each of the four substrips (M,
N, Q, \& R) independently.)  The QR strip is at a declination of $
\delta = +47^\circ \, 34 \arcmin \, 47 \arcsec $, and extends from $8
\hour - 17 \hour$ in right ascension.  The MN strip is at a
declination of $ \delta = +46^\circ \, 22 \arcmin \, 40 \arcsec $, and
extends from $9\hour - 17 \hour$ in right ascension.

As described in more detail in SSG, a single bias level, measured in
the overscan region, was subtracted from each image.  Following this,
the lines of all of the images were combined to create a sky-flat for
the CCD.  The flat did not change significantly over the course of
three years.  After each image was flattened, a cubic polynomial was
fit to the median sky level along the scanning direction and then
subtracted to remove any large temporal fluctuations in the image.
Further corrections of the temporal variations are described in
\S\ref{flattening}.  Finally, the two substrips were calibrated
against each other by requiring the same mean sky level in each pair
of adjacent images.  Each image was further overscan corrected by
using a five piece cubic spline fit to the overscan region.  Following
this, the images were trimmed to remove the bias strip, leaving
372x1598 images.

The imaging data were not taken in perfectly photometric conditions,
and were initially not photometrically calibrated.  However, strip
scans allow one to easily detect the presence of cirrus.  First,
because the strip is a scan both in time and in right ascension, any
temporal variations due to the presence of cirrus is easily recognized
as structured banding within an individual image, due to the
fluctuating brightness of the sky produced by the changing degree of
reflection of the light from San Diego.  Second, the passage of cirrus
dramatically drops the number of objects seen in a given image.  Based
upon the number counts of Postman et al.\ (1996) using the identical
observational system, if the depth were to change by 1 magnitude, the
number of detected objects would change by a factor 3.  In contrast, in the
data which we used, the number of objects never varies by more than
25\% (either RMS or systematically), suggesting that the depth never
varies by more than 0.3 magnitudes.  In fact, the variation in the
depth is certainly even smaller, given that half of the fluctuation
amplitude is due to simple Poisson counting, and that there are
additional fluctuations in the area lost to bright foreground stars.
Using the constancy of the sky brightness and of the number of objects
per field as a measure of atmospheric clarity, we have restricted the
data to the portions of the night which are most likely to be
photometric.  When reduced to the most photometric portions of the
night, the QR substrips consist of 179 400x1600 images in each band
between $8 \hour - 13 \hour$.  The MN substrips consist of 283
400x1600 images in each band.  Twenty-eight of the MN images near $11
\hour $ are unusable due to the passage of clouds.  A total of 868
images were used in the survey.  The total area covered by the 4
individual strips is $17.5\degree^2$.

As a test of our judgement, and as a means to calibrate the zero point
of the Palomar data, we have taken calibration images of many fields
within the strip, under photometric conditions, and find no
significant variation in the zero point of the strip scan data in the
portions of the night to which we have restricted the data. We have
calculated the zero points for the MN and QR strips using calibration
data taken in Johnson $V$ with the Kitt Peak 0.9m telescope in March
of 1994.  The Johnson $V$ zero point of the Kitt Peak data was
calibrated and airmass corrected using 6 independent observations of 8
different standard stars in two separate Landolt (1992) fields, with
an 0.015 magnitude rms in the airmass corrected zero point.  For the
MN strip, we have Johnson $V$ images of 7 different fields throughout
the strip, 3 of which were observed more than once.  These images were
used to calculate $V$ band magnitudes for objects which fell in the MN
strip.  Comparing to the count rate for objects in the SSG data to
their $V$ magnitudes in the KPNO data, we calculate a Johnson $V$ zero
point for the MN strip of $29.36\pm0.04\,$magnitudes per data number,
with no systematic difference between the zero points of the two
halves of the strip.  We will quote Johnson $V$ magnitudes for the
remainder of the paper, for ease of comparison to other LSB samples;
although the data was taken in $V_4$, the color term to correct from
$V_4$ to Johnson $V$ is small, $V-V_4\approx0.04-0.06$, for
$V_4-I_4\approx0.5-1$ (Postman et al.\ 1996).  The RMS scatter among
the 10 calculated zero points of the SSG data agreed with the expected
uncertainty in the mean, based upon the uncertainties in the zero points
calculated for the individual fields, which were in turn based upon the RMS
scatter in the airmass corrected zero point of the KPNO data and in
the zero points calculated for the stars in the KPNO transfer
standards.  The stability of the zero point throughout the MN strip
suggests that the atmosphere was uniformly transparent for the portion
of the night which met our criteria for being most photometric.  The
zero point for the MN strip is also consistent with the conversion
used for the Palomar Distant Cluster Survey (Postman et al.\ 1996),
which uses the same instrumental configuration.

For the QR strip, we have only one calibrated field, giving a zero
point of $29.17\pm0.04$ magnitudes per data number in Johnson $V$ for
the QR strip.  However, rather than use a single measurement to
calibrate the entire QR strip, we will use the zero point of the MN
strip for the QR data as well, and recognize the possibility of an 0.2
magnitude offset between the two data sets. We note that the surface
brightnesses probed in this paper are up to $7\surfb$ fainter than
normal galaxies, and thus an error in the photometric calibration of
less than half of a magnitude will do little to change the
conclusions. Furthermore, given that the number of detected objects
per frame and the sky brightness both vary as smoothly in the QR strip
as in the MN strip, we do not anticipate finding field-to-field
variations in the QR strip zero point which are significantly greater
than those seen in the MN strip.

\subsection{Removal of Foreground Objects}		\label{cleaning}

Producing a map of the background sky requires the removal of all of
the high-surface brightness stars and galaxies which are obstructing
the view of the background.  To identify the intervening objects,
FOCAS was used to identify all regions of connected pixels greater
than 4 sigma above the mean sky, with a detected area of greater than
six pixels after smoothing with FOCAS's built-in filter (Jarvis \&
Tyson 1981, Valdes 1982).  The number of objects found per unit
magnitudes began to deviate from a power law at $V=22$, suggesting
that the FOCAS selection is incomplete beyond this point.

In preparation for cleaning, the region that FOCAS had associated with
each object was extended by an amount derived from the object's
detected area and magnitude.  This ensured that any low-surface
brightness halo, due to either seeing or to a slowly falling galaxy
profile, was included with the central bright regions.  The pixels
within the object were replaced with values drawn from the local sky
histogram, created by sampling within a buffer around the extended
region and clipping at 4 sigma above the local mean.  This cleaning
procedure preserves the large scale distribution of surface
brightness.  Objects which have large areas for their magnitude (or
low magnitudes for their size) are excluded from cleaning to avoid
cleaning any brighter LSBs which FOCAS may have been able to detect
without additional processing.

While replacing objects with the local sky histogram proved to be
effective for objects fainter than roughly 19th magnitude, cleaning
brighter, larger objects, would have invested tremendous amounts of
computational time in a region guaranteed to be useless for detecting
background objects.  Instead, large regions around bright objects were
removed from the survey by creating masks for each image, which were
used throughout processing.  Occasionally, the area of the masked
region was sometimes too small to completely block out the halo of the
masked object.  This problem was most commonly manifested in objects
on the edges of images, and very occasionally in highly elongated
non-boundary objects such as edge-on galaxies and meteor trails.
Cases where the masking was insufficient were easily identified in the
final catalog and removed from consideration.  The two columns on the
left and right edges, and fifteen and thirty rows on bottom and top
edges respectively were masked as well, to eliminate the most common
regions of incomplete FOCAS identification from consideration.
Roughly 39\% of the survey area is lost to the masked regions.

In the final cleaned image, the masked regions were set to the median
sky level of the unmasked regions.  The mask and the cleaned image
were then rebinned by a factor of two to save space.

\subsection{Removing Large Scale Variations}		\label{flattening}

In drift scans, there are low amplitude row-to-row fluctuations due to
temporal variations in the sky brightness ($28.7-28.0\surfb$ (2-4 ADU)
peak-to-peak, when smoothed on scales of $10\arcsec$).  These
fluctuations are constant across a given row, and vary smoothly from
row to row, creating very low level bright and dark ``bands'' across
the strip scan.  The efficiency of the survey is compromised by the
fluctuations; significant low-surface brightness objects that lie in
low points are more likely to be missed, while fainter objects that
lie on the high points will be detected more frequently.  (Although
Monte Carlo simulations can measure the variation in efficiency, they
cannot compensate for the lost survey area.)  We flatten the overall
sky background in each image by: (1) setting the masked regions to
zero, (2) taking the mean of each row of the masked image and of the
mask itself, (3) smoothing each 1-dimensional profile with a boxcar
filter $\approx 2\arcmin$ wide (roughly 13 seconds, temporally), (4)
dividing the smoothed image profile with the smoothed masked profile,
and (5) subtracting the resulting 1-d profile from each column of the
original image.  This produces a significantly flatter image with a
roughly zero sky level.

Correcting for large scale variations reduces the survey's sensitivity
to objects of similar scales.  We have not lost much sensitivity in
the case of this survey however, as the geometry of the strip makes it
unlikely that many objects with scales of $2\arcmin$ or larger could
be detected in the first place.  Because of the difficulty in treating
boundaries in any smoothing problem, the flattening tends to become
inaccurate at the beginning and ending of the image, and adjacent to
regions where the entire width of the image has been masked out.
These regions were obvious in the final smoothed image, and were
eliminated from consideration to avoid false detections.
Occasionally there is residual row-to-row structure left in the image
after flattening, on scales smaller than the flattening scale.  These
are also due to temporal variations on scales smaller than 13s, and
are typically fainter than $28.5\surfb$.  

After removing temporal variations, occasionally there are
diagonal structures which cannot be due to temporal variations, at the
$28\surfb$ level.  The most likely sources of this excess structure
are halos of stars off the field of view and emission from galactic
cirrus.  Our current data are not adequate for addressing these
questions, but repeated, multi-color transit scan observations of a
single region, with small random offsets in declination between scans
and a larger FOV, would help to constrain the origin of any residual
structure.

\subsection{Smoothing and Detection}			\label{smoothing}

The cleaned, rebinned, and flattened images produced by the previous
steps are remarkably featureless.  Any residual flattening
inaccuracies are invisible, buried well below the pixel-to-pixel
variations in the image.  Smoothing the cleaned image reduces the
pixel-to-pixel noise dramatically, uncovering low surface brightness
features.  After replacing the masked regions with the image median,
we smooth the image with a radially symmetric exponential profile
($\propto e^{-r/\alpha}$) with $\alpha=5\arcsec$, which optimizes the
signal-to-noise ratio increase for objects with exponential profiles
identical in size and shape to the smoothing filter.  Smoothing with
the $5\arcsec$ filter reduces the pixel-to-pixel noise\footnote{The
``noise'' of an image was measured throughout this survey as 1.4826
times the width of the second quartile of the pixel histogram.  This
reduces to the standard deviation for a Gaussian distribution.
Because objects contribute light to the sky rather than take it away,
using the second quartile reduces the impact of foreground objects on
the measurement of the width of the pixel histogram.} by roughly a
factor of twelve (to $\approx 0.7-0.9$ ADU).  Smoothing with a larger
smoothing length of $10\arcsec$ reduced the level of fluctuations by
an additional factor of only 1.6, indicating that on the scale of the
smoothing kernel ($8\alpha$) the fluctuations in the images are not
random, due to the very low level fluctuations discussed in \S{flattening}.
The histogram of pixel values for the $10\arcsec$ smoothed
images also show deviations from the expected Poisson distribution.
Because of this indication of residual structure on $40\arcsec$
scales, we chose to use only the single scale length to smooth our
data; with a larger field of view and reduced temporal fluctuations
and scattering, a larger range of smoothing lengths would be more
useful than with this particular data set.

A modified version of FOCAS was used to detect all regions in the
smoothed image greater than $\pm3.5\sigma$ above the image median over
an area of 25 pixels, corresponding to 11.2 arcsec$^2$.  The resulting
FOCAS catalog was restricted to eliminate any object touching a
boundary (the only negative ``dark'' objects found were also boundary
objects) or whose center was within $1.5r_{2} + 10\arcsec$ (where
$r_2$ is the intensity weighted second moment radius) of the left or
right edges, or within $37\arcsec$ of bottom edge or $125\arcsec$ of
the top edge.  The buffer on the upper edges is larger to avoid
detecting objects in the overlap region in more than one image.  With
these selection criteria, 868 objects were detected, giving an average of
1 candidate per each frame.  We expect the total catalog to include
only one or fewer spurious detections due solely to Poisson
fluctuations; this is based upon the number of independent detection
areas in the survey, after correcting the survey area for the masked
areas and the buffered regions along the edges of the images, and upon
the peak height of an exponential profile which is $3.5\sigma$ above
the mean at a radius of $2\arcsec$.  The absence of ``dark'' features
and the low Poisson probability of false positives suggests that
effectively all of the detected features are due to true enhancements
in the the mean sky level (although those may in some cases be due to
other than astronomical sources, as discussed below).

\section{Properties and Classification of Detected Objects}  \label{properties}

A single image was created for each candidate object, consisting of
three adjacent panels (see Figure \ref{lsbfig} for an example).  The
first panel is taken from the original uncleaned, unprocessed image.
The second panel is taken from the cleaned image, and has been
expanded by a factor of two to match the angular scale of the first
panel.  The third panel is taken from the smoothed image in which the
object was detected, also adjusted to the same angular size as the
other two panels.  All three panels are centered on the object.  These
composite images allow easy browsing of the catalog (with much less
stringent disk space requirements than saving all the processed data),
and were used to assign a classification to the objects, if possible.
Classifications were based entirely upon visual inspection of the
images.

The classification was given as combinations of the
categories below, based upon the associated criteria.  The percentages
in the brackets are the percentage of objects that have the category
as a possible classification and the percentage of objects that have
the category as their only possible classification.

(l) Low surface brightness galaxy -- Exhibits smooth, extended emission in the
original image.  [4.1\%,2.4\%]

(c) Cluster -- Shows evidence that the low surface brightness emission
is associated with several compact clumps, either clearly visible or
just below the threshold for detection (see Dalcanton 1996).  The
classification typically reflects one of three cases, corresponding to
clusters at different distances: 1) the light is the low-surface
brightness component of a nearby cluster that can be clearly
identified by large numbers of galaxies in the field; 2) the emission
is centered on a single galaxy with several, barely visible
companions, probably corresponding to a CD galaxy of a more distant
cluster; 3) the emission is centered on a clump of several faint knots
of emission, which are probably the centers of distant cluster
galaxies just below the threshold of detection.  Objects classified as
clusters tend to be near other clumps in the smoothed sky maps.  In
the case of the low redshift clusters, the correlation is due to
detecting pockets of dwarf galaxies and tidal debris throughout the
cluster region.  In more distant clusters, the correlation is most
likely due to the presence of subclustering, or to the high amplitude
of the cluster-cluster correlation function.  This clustering behavior
was clearly seen in the recovery of the $z=0.8$ cluster MS1054 in
Dalcanton (1996), where two or three strong peaks were seen in the
smoothed sky background, centered on regions of truly high galaxy
concentration. [24.3\%,19.9\%]

(t) Tidal features -- Shows evidence that the low surface brightness emission
might be due to interactions between galaxies.  Used if the emission
lies between two close galaxies or is adjacent to a single galaxy that
shows signs of being disturbed, especially if there is no
other evidence that the light might come from galaxies below
the detection limit.  The boundary between this class and the cluster
class is not always distinct. [2.0\%,1.1\%]

(g) Groups of Galaxies -- Tight groups of foreground galaxies without
extended low surface brightness emission.  These
probably entered the catalog through contamination of the sky histogram
of one galaxy with the bright pixels of an adjacent galaxy.  However,
they form a potentially interesting subclass, and may actually contain
true members of the cluster and tidal classes, and thus are not
classified as errors.  [5.4\%,2.8\%]

(?) Uncertainty -- Represents uncertainty in the
classification.

( ) Unclassified -- No evidence for making a classification. Typically
has no strong associated foreground emission, or no compelling reason
to group the object with one class over another. [52.3\%,52.3\%]

(*) Errors -- Cases where the object's inclusion in the catalog was
more a product of the processing than the object's internal
properties.  These cases include incompletely cleaned or incompletely
masked objects, or objects which have only been included because
they were superimposed on either a halo from an object off the field of view
or on regions of inaccurate flattening in the upper and lower edges of
the image. [16.2\%,16.2\%]

Many of the detected objects have more than one possible
classification.  This is an unavoidable consequence of the detection
procedure.  While smoothing the cleaned images increases the
signal-to-noise to the point where faint features can be detected, the
signal-to-noise of the original image, which is used for
classification, remains unchanged.  Thus, in many cases there is no
unambiguous classification, in spite of a significant detection of an
EBL fluctuation.  For example, there are cases where the feature in
the smoothed image does not correspond to any obvious emission in the
foreground image.  There are other cases where, although the detected
region corresponds to one or more high surface brightness features in
the original image, the low-surface brightness component responsible
for the detection is too faint for any morphology to be visible which
might provide a clue as to the proper classification.  Note that
because classification requires more signal-to-noise than detection,
classified objects are even less likely to be spurious.

There are many cases where, in spite of reasonable signal-to-noise,
the classification is ambiguous.  An LSB galaxy with a faint bulge or
with several HII regions is possibly similar in appearance to a
distant cluster with a faint CD galaxy, or even to a tidally disturbed
normal galaxy.  Because of this confusion, when we create a final
catalog of objects which have LSB as their only possible
classification, we are underestimating the true contribution of LSBs.
Because of their ambiguous classification, we systematically exclude
LSBs with high-surface brightness features (i.e. bulges, resolved
stars, HII regions) from our catalog of definite LSBs, leaving us with
a ``pure disk'' subsample of the larger LSB population (e.g.\ see
profiles in Figure \ref{profilefig}).  This is in contrast to visually
selected, angular-diameter limited catalogs of LSBs, such as the
POSSII and UGC catalogs; a substantial fraction of the LSBs in these
catalogs have strong central bulges (for example, see the profiles in
McGaugh \& Bothun 1994).

\section{The LSB Subsample} 		\label{lsbsamp}

For the remainder of the paper, we will concentrate on the 21
objects which were classified as candidate LSBs based upon their
optical morphology.  Of these, 14 were
classified as definite LSBs, and an additional 7 were classified as
probable LSBs.

In order to determine the central surface brightness and scale length
of the LSB candidates, we fit the 21 LSB candidates interactively with
IRAF's ``ellipse'' surface brightness photometry package.  We then
further restricted the LSB subsample to include only the 14 LSB
candidates whose extrapolated central surface brightnesses were
fainter than $23\surfb$ in $V$ (corresponding to roughly $23.5\surfb$
in $B$, using the median LSB color of $B-V\approx0.5$ from McGaugh \&
Bothun (1994) and de Blok et al.\ (1995)); because we did not use any
photometric information in the initial classification, galaxies
classified as potential LSBs were sometimes not dramatically fainter
than the Freeman value, and within the range of surface brightnesses
that can be better constrained with the much larger APM survey (Impey
et al.\ 1996, Sprayberry et al.\ 1997).  The remaining sample spans a
range in central surface brightness of $23\surfb < \mu_0 < 25\surfb$
in $V$ and scale lengths of $2\arcsec<\alpha<7\arcsec$.  While the
profiles are noisy (particularly at large radii) due to the very low
signal-to-noise in the unsmoothed image, they are consistently better
fit by exponential surface brightness profiles than by deVaucouleur's
profiles.

We estimated the uncertainty in the measured central surface
brightness and exponential scale length by performing the identical
ellipse fitting and profile fitting on artificial galaxies with the
measured properties of each of the LSBs.  For galaxies with
$\alpha<5\arcsec$, $\mu_0-\mu_0(true)=-0.03\pm0.06$, and for galaxies
with $\alpha>5\arcsec$, $\mu_0-\mu_0(true) = 0.06\pm0.06$.  For
galaxies with $\mu_0<23.8\surfb$, $[\alpha- \alpha(true)]/\alpha =
-0.006\pm0.09$, and for $\mu_0>23.8\surfb$, $[\alpha-
\alpha(true)]/\alpha = -0.1\pm0.12$.

\subsection{Classification Efficiency} \label{classification}

To measure the number density of LSBs, we must determine the
efficiency with which we can both detect and identify LSBs as a
function of surface brightness and size.  As discussed above
(\S\ref{properties}), because our classification procedure effectively
eliminates LSBs with bulges from the final sample, we need only to
test the recovery of pure disk exponential profile galaxies in order
to measure our efficiency.  While doing so means that we will not be
measuring the efficiency of detecting all LSBs (i.e.\ LSBs with and
without bulges and/or bright HII regions, which are seen to exist in
brighter LSB samples), we know {\it a priori} that our classification
procedure reduces the efficiency of finding such galaxies in our final
sample to zero, given that the surface brightness profiles and images
demonstrate that none of these galaxies with more complicated
morphologies have made it into the final sample.  Thus, the population of
LSB disks with associated bulges and bright HII regions is entirely
unconstrained by this paper, and the measured number density will be
an underestimate of the true LSB population.  Furthermore, because the
galaxies in our sample are all smooth exponential disks with low
ellipticities (with the possible exception of Q-42-1), measuring the
efficiency of the survey requires only that we test the recovery of
similarly smooth, face-on, exponential disks.

To first measure our ability to classify LSBs, we added 100 artificial
LSBs to various positions throughout the original imaging data and
then clipped out the region containing the LSB, creating an image
identical in size to the images which were used for classification in
the actual survey.  The pure exponential face-on disks used as test
LSBs were morphologically indistinguishable from the LSBs in the
sample, which also have pure exponential profiles and low ellipticity.
As a result, the images of the artificial LSBs appeared remarkably
similar to the true LSB images (see Figure \ref{lsbfig}).  We
inspected and classified the artificial LSBs, in the same manner as
the actual survey data.  Figure \ref{classfig}(a) shows the resulting
classifications as a function of exponential scale length and surface
brightness; the triangles were classified as LSBs and the diagonal
lines are lines of constant signal to noise (i.e.\
$\mu_0=\mu_{zp}+2.5\log{\alpha}$, with varying $\mu_{zp}$, and
$\alpha$ measured in arcseconds).  Figure \ref{classfig}(b) shows the
resulting classification efficiency as a function of signal-to-noise
(characterized by $\mu_{zp}$), for $\alpha>2.5\arcsec$ (for scale
lengths smaller than $2.5\arcsec$, the classification ability drops
dramatically, due to confusion between single, low redshift LSBs and
pairs of faint, distant normal galaxies).  Our classification
efficiency is essentially 100\% for $\mu_{zp}<22.2\surfb$, and zero
for $\mu_{zp}>22.95\surfb$.  We approximate the efficiency in Figure
\ref{classfig}(b) analytically as

\begin{equation}			\label{classeqn}
\epsilon_{class}=\cases {
	1					&if $\mu_{zp}\le 22.2$ \cr
	\frac{22.95-\mu_{zp}}{22.95-22.2}	&if $22.2 < \mu_{zp}< 22.95$\cr
	0 					&if $22.95 \le \mu_{zp}$,}
\end{equation}

\ni which has a residual rms of 0.094 for $21.8<\mu_{zp}\le22.95$.

Based upon equation \ref{classeqn}, we make the final restriction on
the subset of LSBs, retaining only those galaxies for which
$\alpha>2.5\arcsec$ and $\mu_{zp}\equiv\mu_0-2.5\log{\alpha}\le22.95$.
This leaves a final LSB catalog of 7 galaxies which have securely
measured classification efficiencies.  The classification images of
these LSBs are shown in Figure \ref{lsbfig} (including also the
cleaned and smoothed images, as described in \S\ref{properties}).
Their radial profiles are shown in Figure \ref{profilefig}, and their
structural properties are listed in Table 1.

\subsection{Detection Efficiency} \label{detection}

In addition to our efficiency in classifying LSBs, we must also
quantify our ability to detect LSB candidates in the first place.
To do so, we used monte carlo simulations to measure the efficiency
with which our automatic detection software identified the smoothed
fluctuations produced by artificial galaxies, as a function of
central surface brightness and exponential scale length.  

First, a single artificial, face-on galaxy with an exponential profile
was added to an image from the survey.  The artificial galaxies were
drawn from a uniform distribution in $\log{\alpha}$ and central
surface brightness, between $1\arcsec<\alpha<125\arcsec$ and
$23<\mu_0<28.25\surfb$, above a fixed signal-to-noise
($\mu_0<2.5\log{\alpha}+26.5\surfb$).  The altered image was processed
in an identical manner to the original survey data.  The resulting
catalog of low surface brightness objects was searched for any
detection that coincided with the artificial galaxy.  This process was
repeated five times for each image in the survey.

The small surface density of true objects ($\approx1.0$ per frame)
implies that confusion and overlapping of low surface brightness
objects is not a source of selection biases in the survey.  Thus, the
Monte Carlo simulations were designed to test the recovery of isolated
galaxies, by requiring that there was at least a $40\arcsec$ buffer
between an artificial galaxy and the nearest candidate LSB.  However,
the detection efficiencies are modulated by any residual large scale
variations in the background sky level, with more objects being
detected in areas where the local sky is higher than average.
Therefore, if too large a buffer is required between the artificial
galaxies and the real objects, then the artificial galaxies will be
systematically biased away from these regions of increased detection
efficiency.  The largest fluctuations occur on scales of
$\approx2\arcmin$, so by limiting the buffer to $40\arcsec$
(substantially smaller than the width of the strip) we have allowed
the artificial galaxies to sample the same sky levels as the real
detections.

Figure \ref{detectionfig}(a) shows the resulting detection efficiency
for our survey, as a function of surface brightness and disk scale
length.  Superimposed on the contours of constant efficiency are all
the LSBs with $\mu_0\!>\!23\surfb$, as well as the regions of 100\% and
$>\,0$\% classification efficiency.  Within the region where LSBs can be
reliably classified, the detection efficiency is well behaved, as can
be seen in Figure \ref{detectionfig}(b), where we have plotted the
detection efficiency as a function of scale length, for different
central surface brightnesses.  The distribution is
reasonably well fit by a Gaussian, leading us to use

\begin{equation}			\label{detectioneqn}
\epsilon_{det} = 0.37 \exp{\left[-\log^2{(\alpha/6.3\arcsec)}/0.15\right]}
\end{equation}

\ni as a reasonable approximation of the detection efficiency for the
final LSB subsample.  The residual rms of this approximation is 0.056.
The maximum detection efficiency is roughly 35\%, which results largely
from the loss in area behind masks, behind other stars and galaxies,
and around the edges of the images.  The detection efficiency peaks
close to the scale length of the $5\arcsec$ exponential smoothing
filter used in processing.  The contours of constant detection efficiency
suggest that LSBs with $\mu_0 = 26\,V\surfb$ may be in the full
catalog, but be currently unclassified or misclassified as clusters.

We have not to attempted to model edge-on LSBs.  Including edge-on
LSBs could have easily introduced more uncertainty than it would have
removed; we know very little about the vertical scale heights and
opacities of LSB disks, particularly at the extreme surface
brightnesses explored by this survey.  Furthermore, none of our
detected LSBs is substantially elongated, and thus the detection
efficiency for edge-on LSBs is not an essential ingredient for
recovering the LSB number density of the smooth face-on LSBs in our
sample.  However, since we are strongly biased against finding edge-on
LSBs, due to our choice of a circularly symmetric smoothing filter, we
may well be underestimating the number density of LSBs by effectively
restricting our sample to include mostly face-on galaxies.

\subsection{Distances}				\label{distances}

Follow-up spectroscopy of the final LSB subsample was
carried out in May of 1996 using the CryoCam on the Mayall 4m at Kitt
Peak National Observatories.  The 650 grism (400 line/mm) was used in
first order with a $3.2\arcsec$ slit, giving $\sim\!13$\AA\ resolution
(FWHM) between 3500\AA\ and 7000\AA, with an average of 3\AA\ per
pixel.  The slit was oriented along the major axis of the LSB, using
the Richley prisms to correct for the differential refraction of the
atmosphere.  Exposure times ranged from 2700s to 12,000s giving a
signal-to-noise of 10-40 per resolution element longwards of the the
4000\AA\ break, with a mean signal-to-noise of 15-20.  

The spectra of the LSBs were overscan corrected, dark
subtracted, flat-fielded using domeflats, illumination corrected with
twilight flats, and then extracted, sky subtracted, and wavelength
calibrated using He+Ne+Ar spectra taken adjacent to, or sandwiched
between, each of the LSB spectra.  The focus of the
spectrograph was a strong function of wavelength, and degrades by
nearly a factor of two by 3600\AA.  

The resulting LSB spectra were initially cross-correlated with a
template spiral galaxy spectrum using IRAF's ``rvsao'' package, and
searched for emission lines.  The cross-correlation was limited to
between 3700\AA\ and 5000\AA\, to isolate the 4000\AA\ break, the
Balmer lines, and the strong G-band.  Since all galaxies were at very
low redshift, the correlation included nearly the same region of the
rest-frame spectra for all galaxies.  Using the resulting redshifts,
the spectra were doppler corrected back to zero redshift.  Then, to
refine the redshift measurement, we created an improved correlation
template for each galaxy, by coadding all the other LSB spectra,
excluding the one galaxy spectra to be cross-correlated and the
spectra for R-26-1, which has low signal-to-noise and thus no
convincing redshift.  The new templates were then cross-correlated
with the correct galaxy spectra and a new redshift derived.  The
process was repeated a second time using templates derived from the
refined redshifts.  On the final cross-correlation, no redshift
changed by more than $450\kms$, and all correlation coefficients had
$4.4<R<7.6$, with the exception of M-232-1, which had $R=3.7$, and
R-26-1, for which we have no convincing redshift.  The emission line
redshifts for R-27-1 and Q-129-2 were within $400\kms$ of the cross
correlation absorption line redshifts.  We take this velocity
difference to be representative of the $1\sigma$ uncertainty in the
individual redshifts, although the formal error from the
cross-correlation is smaller ($\lta200\kms$).  Similar changes in the
velocity were produced by changing the wavelength range and filter
parameters used for the cross-correlation, again suggesting that the
redshifts given in Table 1 are accurate only to several $100\kms$.  In
all cases, however, the cross-correlation functions do not have
additional peaks which could be considered as alternative plausible
redshifts.  The redshift for M-232-1 is more uncertain, due to the
lower signal-to-noise and as can be seen from the lower correlation
coefficient.  By eye, we find a limit of $\pm1000\kms$ on the error in
the redshift.  The final redshifts are listed in Table 1, and the
spectra are shown in Figure \ref{spectrafig}, along with the template
spectra used for cross-correlation, all shifted into the rest frame;
because of the low redshifts, the shift was only $\sim100$\AA.  For
the purposes of analysis, we assign a redshift to R-26-1 by assuming
it has the mean physical scale length of galaxies in the sample
($h=2.9\kpc$), giving an assumed recessional velocity of $11600\kms$.
If R-26-1 were to have the smallest or largest physical scale length
of the sample, it would have recessional velocities of $6800\kms$ or
$14600\kms$ respectively.

The LSBs have recessional velocities of 4000-9000$\kms$, and disk
scale lengths of $1.7-3.6\,{\rm h}_{50}^{-1}\kpc$ (Table 1, excluding
R-26-1, due to its uncertain redshift), comparable to the disk scale
lengths seen in angular diameter limited field surveys of brighter
LSBs (McGaugh \& Bothun 1994, de Blok et al 1995, de Jong 1996) and of
normal galaxies.  The LSBs in the final subsample are also at
comparable distances to normal galaxies catalogued in large local
surveys.  The similarity can be seen most readily in Figure
\ref{piefig}, where we have plotted a ``pie'' redshift-position
diagram for the final LSB sample (solid triangles) and for the
galaxies in the ZCAT catalog of galaxies (open circles).  The LSBs
also have scale lengths which are similar to normal galaxies.  Both
the LSB and normal galaxy samples will be biased towards finding
physically large galaxies, which have the largest probability of being
identified in angular diameter limited surveys (see Appendix B).

Figure \ref{piefig} shows that there is no strong tendency for the
LSBs in our sample to be physically associated with normal galaxies.
Between $3000\kms$ and $10000\kms$, 11 out of 23 ZCAT galaxies are
in close associations, whereas 2 in 6 of the LSBs has a near
neighbor.  Because of the small numbers of LSBs and the non-uniform
selection criteria for the ZCAT galaxies, our sample does not provide
definitive evidence for the low correlation between LSBs and normal
galaxies.  However, it is suggestive of the general trend seen in brighter
LSB samples for the small-scale correlation of LSBs to be weak
compared to normal galaxies (Bothun et al.\ 1993, Mo et al.\ 1995).

The follow-up, spectroscopic observations also provide validation for
our classification procedure.  All of the candidate LSBs in the final
sample which we observed spectroscopically have characteristic galaxy
spectra, smooth light profiles perpendicular to the dispersion
direction, and low redshifts, proving that the final sample includes
only the nearby LSBs which the survey was designed to find.  While our
classification procedure may have missed LSBs (i.e. false negatives),
our strict selection criteria (\S\ref{detection} \& \ref{classification})
successfully insured that we did not include any non-LSBs in our final
candidate list (i.e. false positives).


\section{The Number Density of LSBs}		\label{numberden}

Using the classification and detection efficiencies from equations
\ref{classeqn} \& \ref{detectioneqn} we can correct for the area that
was effectively lost from the survey and calculate the surface density
of LSBs with $23<\mu_0<25\,V\surfb$.  The total efficiency
$\epsilon(\mu_0,\alpha)=\epsilon_{class}(\mu_0,\alpha) \times
\epsilon_{det}(\mu_0,\alpha)$ is a measure of the fraction of the
total $17.48\,{\rm degree^2}$ survey area which was actually
accessible for detecting an LSB with central surface brightness
$\mu_0$ and exponential disk scale length $\alpha$.  Using the
formalism developed in Appendix A, the surface
density of LSBs in our survey with $23<\mu_0<25\surfb$ and
$\alpha>2.5\arcsec$ is:

\begin{equation}			\label{surfdeneqn}
\Sigma(23<\mu_0<25\surfb,\alpha>2.5\arcsec) 
		= 4.1^{+2.6}_{-2.1}\, {\rm galaxies / degree^2},
\end{equation}

\ni where the errors enclose 90\% confidence.  The uncertainty is
dominated by the Poisson probability for 7 galaxies and a detailed
calculation of the confidence interval is provided in Appendix A.
While a naive calculation of the surface density (7 galaxies in 17.5
square degrees) gives a smller surface density, inspection of Figure
\ref{detectionfig}(A) for $\alpha>2.5\arcsec$ shows that the detection
and classification efficiencies are small over much of region, leading
to a large correction.  As shown in Appendix B, the surface density
of galaxies on the sky increases as $\alpha^{-4}$ towards small
angular sizes, and thus most of the correction applied to the
measured surface density is a correction for the lost classification
efficiency of small galaxies with $\mu_0>24\surfb$.  Furthermore,
because of the steep increase in the number of galaxies with small
apparent scale lengths, the surface density is a strong function
of the exact lower limit on $\alpha$, and thus LSB surveys with
different angular selection criteria will derive very different
surface densities.

Although the surface density may at first seem small, it is in fact
significant, given that the large scale lengths to which the sample
was restricted effectively confines the survey to a small, nearby
volume.  The low surface density also explains why there is a
consistent impression that LSBs are rare.  First, one would have to
observe a square degree of the sky to be reasonable certain of finding
one large LSB.  Second, the LSB would appear much smaller than a
normal galaxy of comparable scale length, because it would only be
visible out to roughly one scale length before its surface brightness
drops below the pixel-to-pixel variations in the sky; normal galaxies
are visible out to several scale lengths and thus are more noticible
(Disney \& Phillips 1983).

Using the distances for the final LSB subsample, the efficiency for
detection and classification and the selection criteria for the final
sample, we can calculate the effective volume in which each LSB could
be detected.  An LSB at a distance $D_i$, with angular scale length
$\alpha_i$ and central surface brightness $\mu_i$ could have been detected
out to a distance of

\begin{equation}				\label{Dmax}
D_{max_i}=D_i \times \cases {
	\frac{\alpha_i}{2.5\arcsec} &if $\mu_i < 22.95+2.5\log{2.5\arcsec}$ \cr
	\frac{\alpha_i}{10^{0.4(\mu_i-22.95)}}
				    &if $\mu_i \ge 22.95+2.5\log{2.5\arcsec}$}.
\end{equation}

\ni The effective volume of the survey is different for each galaxy,
because the detection efficiencies are a strong function of the
angular scale length and central surface brightness of the individual
galaxies.

Using $D_{max}$, the volume that the $i$th galaxy could be detected within
is

\begin{eqnarray}				\label{V}
V_i &=& \frac{A}{\rm radians}\,
	\int_0^{{D_{max}}_i}
	      \epsilon(\mu_i,\alpha_i\left[\frac{D_i}{D^\prime}\right])\,
	      {D^\prime}^2\,{\rm d}D^\prime \\
    &=& \frac{A}{\rm radians}\,
	\left[\alpha_i D_i\right]^3
	\int_{\frac{\alpha_i D_i}{D_{max_i}}}^\infty
	      \frac{\epsilon(\mu_i,\alpha^\prime)}{{\alpha^\prime}^4}\,
	      {\rm d}\alpha^\prime,
\end{eqnarray}

\ni including the change in detection efficiency with the apparent
angular size; because the efficiency changes with distance, the survey
volume associated with a given galaxy is not a simple conical section,
but tapers off as $D$ approaches $D_{max}$.  There is no minimum
distance at which an LSB can be detected, because we imposed no upper
limit on the angular size of galaxies within our survey.  There is an
implicit upper limit set by the width of the survey fields, but the
limit is so much larger than the angular scales of the galaxies in the
survey that the effective minimum distance is zero.

With equations \ref{Dmax} and \ref{V}, the total number density of
LSBs with $23<\mu_0<25\surfb$ and intrinsic scale lengths $h$ between
$1.7\,h_{50}^{-1}\kpc$ and $3.6\,h_{50}^{-1}\kpc$ is

\begin{eqnarray}				\label{numden}
{\cal N}(23<\mu_0<25\surfb,1.7<h<3.6h_{50}^{-1}\kpc)
	&=& \sum_{i=1}^n \frac{1}{V_i} \cr
	&=& 0.01^{+0.006}_{-0.005} \, {\rm galaxies / h_{50}^3 Mpc^3},
\end{eqnarray}

\ni with 90\% confidence, assuming the probability distributions
derived in Appendix A.  As with the measurement of the surface
density, the uncertainty is entirely dominated by the uncertainty due
to the small number of galaxies in the final LSB subsample.  The value
of ${\cal N}$ changes negligibly other reasonable redshifts for
R-26-1 are used.

In order to compare the relative contribution of LSBs and normal
galaxies, Figure \ref{numdenfig} shows the integrated number density
of LSBs with central surface brightness fainter than $23\surfb$ in
$V$, as a function of limiting surface brightness.  Superimposed on
the integrated LSB number density are comparable measurements of the
integrated number densities of normal galaxies, for several different
determinations of the local luminosity function.  To make
``comparable'' integrated number densities of normal galaxies we have
restricted the integration to either (1) the range of absolute
magnitude covered by the LSBs in our sample
($-16.1>M_V+5\log{h_{50}}>-18.6$) or (2) the range of intrinsic
exponential disk scale lengths of our sample
($1.7-3.6\,h_{50}^{-1}\kpc$) (assuming that a normal galaxy's scale
length can be determined from its luminosity, if it has a Freeman
central surface brightness of $21.7\surfb$ in $B$ and $\langle
B-V\rangle\sim0.5$).  The corresponding number densities are labelled
in Figure \ref{numdenfig} as ``same M'' and ``same $\alpha$'',
respectively.

As Figure \ref{numdenfig} shows, the integrated number density of LSBs
with central surface brightness fainter than $23\surfb$ in $V$ is {\it
greater} than the number density of normal galaxies with either
similar luminosities or similar scale lengths.  We consider this result
to be extremely robust.  The errors are dominated by the
well-understood Poisson statistics of the small number of LSBs, and
thus, only an egregious, systematic error in the measured distances or
in the integrated efficiencies (eqn \ref{V}) could produce a large
enough effect to change our conclusions.  In spite of the small
numbers of LSBs in the final sample, the large number density of LSBs
would persist even if significant fractions of the sample were to be
removed.  For example, even between $23\surfb$ and $24\surfb$, the
number density of LSBs remains comparable or greater than normal
galaxies.  Our conclusions hold as well if the confidence intervals
are increased to 99\%.

We have also considered the degree to which photometric errors could
effect our measurement of ${\cal N}$, by examining $1/V_i$ as a
function of central surface brightness $\mu_0$.  Brighter than
$\mu_0=24\surfb$, the effect of photometric errors is very small, such
that an error of 1 magnitude in the zero point changes the derived
number density by less than a factor of two, due to variations in the
integral over the efficiency.  The effect is stronger for fainter
central surface brightnesses, where the overall efficiency is lower
and more sensitive to $\mu_0$.  Even so, the required zero point error
would need to be larger than the observations admit.  For example, to
reduce the number density for $\mu_0>24\surfb$ by a factor of 10
(which would still leave ${\cal N}$ comparable to the normal galaxy
number density), the zero point would have to be systematically 0.9
magnitudes too faint in the fields containing the faintest LSBs.  From
our photometric calibration, we believe that our mean zero point is
good to at least 0.1 magnitudes, so there would have to be a temporary
fluctuation in the zero point of the strip scan while the LSBs were
passing overhead.  Such a shift would have produced nearly a factor of
3 drop in the mean number of objects per field, based upon the number
counts of Postman et al.\ (1996) and would probably have been
accompanied by a brightening of the sky level.  Neither of these
signatures is seen, given that the data was originally restricted to
the parts of the strip scan which were stable in both sky brightness
and number of detected objects.  The number of objects never varies by
more than 25\% in either the RMS or the mean, suggesting that the zero
point does not fluctuate by more than 0.25 magnitudes, based upon the
number counts.  This is a strong upper limit, given that simple
Poisson statistics contributes 10\% to the RMS component and that
the lost detection area due to bright stars probably makes a similar
contribution.  The maximum zero point uncertainty suggests that, for
$\mu_0>24\surfb$, the most we could have possibly overestimated an
individual galaxy's contribution to the number density is a factor of
2, which is still contained within the 90\% confidence interval.
Thus, the total measurement of ${\cal N}$ could be high by at most a
factor of 2, and only in the unlikely event that every LSB in the
sample had the largest possible zero point error, in exactly the same
direction.

We have reasons to believe that, if anything, our measurement
of ${\cal N}$ is an underestimate.  First, our choice of a circularly
symmetric smoothing filter greatly reduces our sensitivity to edge-on
LSBs, and thus we are only sensitive to the fraction of galaxies which
are seen relatively close to face-on.  This limitation is evident
in the fairly circular morphologies of all of the galaxies found
in the survey.  Second, we have been extremely
conservative in our decision to classify detections as possible LSBs.
We have consciously excluded possible LSBs which we feel cannot be
adequately distinguished from clusters or tidal extensions off bright
galaxies.  Thus LSBs with small HII regions or central bulges are not
included in our sample, and we are restricted to finding only LSBs
which have classical Im morphologies.  There is a tendency for higher
luminosity, large scale length LSBs to be earlier Hubble types
(McGaugh, private communication), and thus we are less sensitive to
the most luminous LSBs.  This bias will lead us to underestimate
the luminosity density even more severely than we are underestimating
the number density.  Note also that the galaxies which we have identified
have very regular, smooth morphologies, allowing us to accurately
simulate their detection and classification efficiencies.

In Figure \ref{numdenfig}, the overdensity of LSBs is smaller when
compared to the AutoFib survey luminosity function (Ellis et al.\
1996), than when compared to the Stromlo-APM (Loveday et al.\ 1992) or
DARS (Peterson et al.\ 1986) surveys, which both have limiting
magnitudes of $b_j\approx17$.  However, the AutoFib survey combines
the shallower DARS survey with much deeper pencil beam surveys which
reach limiting isophotal magnitudes of $b_j\approx24$.  As discussed
by McGaugh (1994), this implies that the AutoFib survey has a fainter
surface brightness limit than the shallower Stromlo-APM or DARS
surveys.  Thus, the AutoFib survey is likely to have cataloged far
more intrinsically low surface brightness galaxies than the shallow
nearby surveys, increasing the integrated number density implied by
the AutoFib luminosity function over what is measured in shallower
surveys; this is clear from the integrated number densities plotted in
Figure \ref{numdenfig}.  (Note also that the AutoFib luminosity
function has the steepest faint-end slope of the three surveys.)  The
different surface brightness ranges of the surveys, therefore, changes
the apparent relative importance of the LSB and normal galaxy
population.

The integrated luminosity density of LSB galaxies is presented in
Figure \ref{lumdenfig}, along with the integrated luminosity density
in normal galaxies.  When restricted to the same range in absolute
magnitude, normal galaxies and LSB with $\mu_0>23\surfb$ and
$1.7-3.6\,h_{50}^{-1}\kpc$ make a comparable contribution to the total
luminosity density of the universe.  However, when restricted to the
same range in scale length, the higher luminosity, high surface
brightness disks dominate the luminosity density by factors of 2-6
over the LSB component, as might be expected by the drastic difference
in intrinsic luminosity.  There is also is very little evidence for a
major contribution to the luminosity density from galaxies whose disks
are fainter than $\mu_0=24\surfb$ in $V$; even if there is possible
evidence for a rise in number density with decreasing surface
brightness in Figure \ref{numdenfig}, it is not enough to compensate
for the increasing faintness of the disk.  We should note however,
that these results apply only for a limited range of galaxy
properties, and should not be applied more generally.  
Because of our bias against LSBs with bulges, we may be systematically
excluding the most luminous galaxies from our sample, and thus
our conclusions only hold for the range of disk scale lengths and
absolute magnitudes covered by our sample.  Our calculated
contribution of LSBs with $\mu_0>23\surfb$ to the luminosity denisity
is comparable to that calculated in the brighter APM LSB
survey (Sprayberry et al.\ 1997).

We may compare the measured relative number densities and luminosity
densities for LSBs and normal galaxies to those which are predicted by
the formalism presented in Dalcanton et al.\ (1997).  We do so by
integrating the predicted galaxy number density for central surface
brightnesses between $23.5\surfb$ and $25.5\surfb$ in $B$ (assuming
$\langle B-V \rangle \sim 0.5$ for the LSBs in our sample), and
intrinsic disk scale lengths between $1.7-3.6\,h_{50}^{-1}\kpc$, and
comparing it to the integrated predicted number density for comparable
normal galaxies, with ``comparable'' defined as in Figure
\ref{numdenfig}, and assuming $20<\mu_0<22.5\surfb$ for the galaxies
which are typically used to determine the local luminosity function;
this limit is most appropriate for the DARS and Stromlo-APM surveys.
We make a similar calculation of the luminosity density.  For the
theoretical models, we have assumed the values of
the values of $\Upsilon=3\msun/\lsun$ for the mass-to-light ratio
of baryons, $F=0.05$ for the baryonic mass fraction, and $M_*=10^{12}\msun$,
and $\alpha_{lum}=-1.5$ for the Schechter function describing the
distribution of galaxy masses.

We find that the predicted ratio of the number density in LSBs to the
number density in normal galaxies with similar scale lengths is 2;
this ratio varies somewhat with variations in the limiting surface
brightness for normal galaxy surveys between $22.0\surfb$ and
$23\surfb$.  The predicted ratio is therefore in agreement with what
is measured in Figure \ref{numdenfig}.  Restricting the normal galaxy
sample to have similar absolute magnitudes to the LSB sample, the
predicted ratio becomes 0.3.  The agreement between the measured and
predicted relative number density is not as good in this case, with
there being even more LSBs measured than predicted.  However,
comparing the same range of absolute magnitude is much more sensitive
to the faint-end slope of the luminosity function at very faint
magnitudes.  Because of the rapid rise in number density with
decreasing luminosity, the integrated number density depends strongly
on the number density of the faintest galaxies
($M_V+5\log{h_{50}}\sim-16$ in this case); unfortunately, this
limiting absolute magnitude is where the local luminosity function is
most poorly determined, and may be extremely incomplete (Driver \&
Phillipps 1996).  The discrepancy may also result from the parameters
of the theoretical model.  We have assumed that the theoretical mass
function of galaxies has a steep faint-end slope of $-1.5$, which
greatly inflates the number of low-mass galaxies relative to high-mass
galaxies; this accounts for the large difference seen in the predicted
ratio of number densities when the normal galaxies are restricted to
similar absolute magnitudes rather than similar scale lengths.  We
have also assumed that a galaxy's mass is uncorrelated with its
angular momentum, whereas in the BBKS formalism (Bardeen et al.\
1996), there should be a correlation wherein low mass galaxies tend
have higher angular momenta (Catelan \& Theuns 1996).  This
correlation would increase the expected ratio of LBSs to normal
galaxies in a fixed magnitude range.

We see a similar level of agreement between the predicted and measured
relative luminosity densities.  The predicted ratio of luminosity
densities of normal galaxies to LSBs with the same range of scale
lengths is $\sim\,6$, which also agrees reasonably well with the
observations (Figure \ref{lumdenfig}.  When restricted to the same
range of absolute magnitude the predicted ratio of normal galaxy
luminosity density to LSB luminosity density is 4, whereas they are
observed to be comparable.  This suggests that there is more
luminosity density in LSBs than predicted.  Again, uncertainty in
the measured faint-end of the local luminosity function, or a
correlation between mass and angular momentum could bring these into
agreement.

\section{Conclusions}				\label{conclusions}

The importance of the low surface brightness galaxy population has
been an outstanding question in astronomy.  Claims have ranged from
LSBs being a negligible component of the local universe (Roukema \&
Peterson 1994) to outnumbering $L>0.1L_*$ galaxies by orders of
magnitude (${\cal N}=1\Mpc^{-3}$ for scale lengths greater than
$0.8\kpc$; Schwartzenberg et al.\ 1995).  To address this recent
debate, we have presented the first well-quantified measurement of the
absolute number density and luminosity density of low surface
brightness galaxies, using a large area CCD survey with well
understood selection criteria, quantifiable detection efficiencies and
follow-up spectroscopy of all LSB candidates.  Our survey has selected
LSBs which are comparable to the galaxies in the NGC catalog, in both
size and distance, allowing a direct, meaningful comparison with local
field galaxy surveys.

We find that the number density of LSB galaxies with $23 < \mu_0 <
25\,V\surfb$ is indeed greater than the number density of normal
galaxies, when the normal galaxies are restricted to the same range in
either absolute magnitude ($-16.1>M_V+5\log{h_{50}}>-18.6$) or in exponential
disk scale length ($1.7-3.6\,h_{50}^{-1}\kpc$).  The exact ratio
depends upon which local luminosity function is used for comparison,
but LSBs typically outnumber comparable normal galaxies by factors
of 2 or more.  The measured ratios are formally consistent with
theoretical expectations based on assuming that
high angular momentum and/or low mass protogalaxies collapse to
form low surface brightness galaxies (Dalcanton et al. 1997).

However, we also find that LSBs with central surface brightnesses
fainter than $23 \,V\surfb$ and with scale lengths between
$1.4-3.6\,h_{50}^{-1}\kpc$ make a less significant contribution
to the luminosity density of the universe.  In our survey, galaxies
fainter than $24\,V\surfb$ make almost no contribution to the
luminosity density, and galaxies fainter than $23 \,V\surfb$
contribute 15-50\% of the luminosity density contributed by
comparable normal galaxies.  Thus, a significant (but not
overwhelming) fraction of the stars, and perhaps the mass,
of the universe is in LSBs.

The inferred number densities and luminosities are likely lower limits
as the survey may have excluded significant number of LSBs due to
selection effects.  By design, our survey does not include galaxies
which are intermediate in surface brightness between $23 \,V\surfb$
and the characteristic Freeman surface brightness; these galaxies are
intrinsically more luminous, and thus could make a much larger
contribution to the luminosity density.  Furthermore, we have
restricted our selection to Sm/Im morphologies, whereas the brightest,
largest scale length LSBs tend to have both bulges and spiral arms
(McGaugh; private communication).  This bias against earlier type
morphologies may well have lead us to underestimate the LSB luminosity
density, and, less severely, the LSB number density.  Our survey also
includes a strong bias against edge-on LSBs (due to using a circularly
symmetric filter in the detection algorithm), which may again lead us
to underestimate the LSB density.

\bigskip
\centerline{Acknowledgements}
\medskip

Tony Tyson and Ian Smail are warmly thanked for advice on the
photometric calibration, as is Lori Lubin for her assistance with data
acquisition at the KPNO 0.9m and Jim DeVeny for his help with the
CryoCam at the KPNO 4m.  Tony also generously donated huge quantities
of disk space, without which this project could not possibly have been
done.  The referee is also thanked for very helpful commentary and
suggestions.  JJD gratefully acknowledges useful discussions with
Rebecca Bernstein and Dan Rosenthal.  Support for JJD was provided by
NASA through Hubble Fellowship grant \#2-6649 awarded by the Space
Telescope Science Institute, which is operated by the Association of
Universities for Research in Astronomy, Inc., for NASA under contract
NAS 5-26555.  Partial support for DPS provided by National Science
Foundation grant AST-95-09919.  MS was supported by NSF grants
AST91-08834 and AST94-15574.

\appendix
\medskip
\centerline{\bf Appendix A}
\medskip
\centerline{Probability Distributions for $\Sigma$ and ${\cal N}$ for
Small Numbers of Galaxies}
\bigskip

Suppose we have a sample of $n$ galaxies, the $i$th which has angular
scale length $\alpha_i$ and central surface brightness $\mu_i$.
Assume that there is a unique overall detection efficiency
$\epsilon(\alpha,\mu)$ associated with each scale length and surface
brightness, and that there is a fixed rms uncertainty in $\epsilon$
of $\sigma_\epsilon$.  

Assuming that the underlying galaxy distribution in surface brightness
and apparent scale length is $f(\mu)f(\alpha)$, the total number
$\overline{N}$ of galaxies expected in a survey of area $A$ with
central surface brightnesses between $\mu_1$ and $\mu_2$ and
apparent angular scale lengths greater than $\alpha_1$ is:

\begin{equation}			\label{nbarappendix}
\overline{N} = \int^{\mu_2}_{\mu_1} \int^\infty_{\alpha_1}
		f(\mu)f(\alpha) \, \epsilon(\alpha,\mu)\,d\mu\,d\alpha.
\end{equation}

\ni If the survey had perfect efficiency, $\epsilon \equiv 1$, the survey
would recover

\begin{equation}
\overline{N}_0 = \int^{\mu_2}_{\mu_1} \int^\infty_{\alpha_1}
		f(\mu)f(\alpha) \,d\mu\,d\alpha
\end{equation}

\ni galaxies.  Thus, the total survey efficiency is $\overline{N} /
\overline{N}_0$.  Suppose then, that an individual survey measures
$n$ galaxies over the survey area.  The surface density becomes

\begin{equation}			\label{sigappendix}
\Sigma  = n \, \frac{\overline{N}}{\overline{N}_0} \, \frac{1}{A}.
\end{equation}

As shown in Appendix B, for local volumes where cosmological curvature
in negligible, $f(\alpha)=\alpha^{-4}$ for any distribution of intrinsic
galaxy scale lengths.  We have assumed that $\mu$ and $\alpha$ are
independent variables, and have the freedom to choose any distribution
of $f(\mu)$.  We take $f(\mu)$ to be constant to first order, as
is consistent with the data presented in Figure \ref{numdenfig}.

To calculate the probability distribution of the surface density
$\Sigma$, we first consider the effect of having a small number of
galaxies in the sample.  We calculate this Poisson uncertainty by
adopting Bayes Theorem.  We assume that, given a single trial which
measured $n$ events, the probability that infinite
number of similar trials would measure a mean of $\overline{n}$ events
is $p_n(\overline{n} | n) \propto p(n | \overline{n})
p(\overline{n})$.  Because we have no prior reason to prefer any value
of $\overline{n}$, we assume that the prior probability
$p(\overline{n})$ is uniform, and thus that $p_n(\overline{n} | n)
\propto p(n | \overline{n})$, which is simply the Poisson probability
of observing $n$ events for a system which would give a mean of
$\overline{n}$ events.

There is also a contribution to the uncertainty in $\Sigma$ from the
uncertainty in the efficiency $\epsilon$.  However, the integral in
equation \ref{sigappendix} averages out much of the uncertainty in the
efficiency $\epsilon$.  For the particular example of our survey,
while the pointwise estimation of $\epsilon_{class}$ and
$\epsilon_{detect}$ has an rms uncertainty of roughly $0.1$, the
integral over the efficiency in equation \ref{nbarappendix} is
much more tightly constrained.  The uncertainty contributed to the
integral by $\epsilon_{class}$ is negligible ($\sim1$\%), and the
uncertainty contributed by $\epsilon_{detect}$ is likewise small
($<\!10$\%) for central surface brightnesses fainter than
$23.5\surfb$.  At brighter surface brightnesses, our approximation for
$\epsilon_{detect}$ overestimates the integral volume by $\sim50$\%.
However, we only have one galaxy in this surface brightness interval,
and the direction of the error is to underestimate the number density,
thus we expect this to contribute very little uncertainty to the total
number density.  Given that the Poisson uncertainty is of the order of
$40\%$ for our sample, we will neglect the contribution of
$p_{\epsilon_i}(\epsilon)$ to the uncertainty in $\Sigma$.  Thus, the
final probability distribution for the surface density $\Sigma$ is

\begin{equation}
p(\Sigma) = p_n(\overline{n} | n=A\Sigma\frac{\overline{N_0}}{\overline{N}})
		\, A\frac{\overline{N_0}}{\overline{N}}.
\end{equation}

We may also solve for the effective number density of the survey,
given the distance $D_i$ of each galaxy, and the maximum distance at
which each galaxy could be detected, $D_{max_i}$ (eqn.\ \ref{Dmax}).
The effective volume associated with each galaxy, $V_i$, is an
integral over distance, accounting for the changing efficiency with
changing angular size:

\begin{eqnarray}				\label{Vappendix}
V_i &=& \frac{A}{\rm radians}\,
	\int_0^{{D_{max}}_i}
	      \epsilon(\mu_i,\alpha_i\left[\frac{D_i}{D^\prime}\right])\,
	      {D^\prime}^2\,{\rm d}D^\prime \\
    &=& \frac{A}{\rm radians}\,
	\left[\alpha_i D_i\right]^3
	\int_{\frac{\alpha_i D_i}{D_{max_i}}}^\infty
	      \frac{\epsilon(\mu_i,\alpha^\prime)}{{\alpha^\prime}^4}\,
	      {\rm d}\alpha^\prime,
\end{eqnarray}

\ni assuming that the galaxies are close enough that $(1+z)^4$
cosmological dimming of $\mu$ can be neglected.  With the exception
of the integral over surface brightness, this integral is
identical to the one equation \ref{nbarappendix}.

The uncertainties in each volume term $V_i$ are dominated by the
Poisson uncertainties which also dominate the calculation of $\Sigma$
(here, $n=1$).  The other possible contributions are from the
uncertainty in the measurement of $\alpha_i$, which is $\sim10$\%, and
in $D_i$, which is less than $5\%$.  These are also negligible compared
to the Poisson uncertainties; in the limit of Gaussian statistics,
including a 10\% uncertainty from the detection efficiency and a 10\%
efficiency from the measurement of $\alpha_i$ would increase a 40\%
Poisson uncertainty to only 42\%.

The number density, ${\cal N}$, measured for the sample of $n$ galaxies is

\begin{equation}
{\cal N} = \sum_{i=1}^n \frac{1}{V_i}.
\end{equation}

\ni With our assumption that Poisson statistics dominate uncertainties
in $V_i$, the probability for the individual terms ${\cal N}_i=1/V_i$ is
$p_{{\cal N}_i}=p_n({\cal N}_i V_i | 1) V_i$.  The final probability
distribution of ${\cal N}$ is then

\begin{equation}					\label{pNappendix}
p_{\cal N}({\cal N}) = A\,
	\int^\infty_0 p_{{\cal N}_1}({\cal N}^\prime_1)
	\int^\infty_0 p_{{\cal N}_2}({\cal N}^\prime_2)
	\int^\infty_0 \cdots
	\int^\infty_0 p_{{\cal N}_{n-1}}({\cal N}^\prime_{n-1})
		      p_{{\cal N}_n}({\cal N} - \sum_{i=1}^{n-1} {\cal N}_i)
		      d{\cal N}^\prime_1d{\cal N}^\prime_2 \cdots
			d{\cal N}^\prime_{n-1},
\end{equation}

\ni which can be solved with Monte Carlo integration by repeatedly
drawing samples of galaxies from each of the $p_{{\cal N}_i}({\cal N}_i)$.

\appendix
\medskip
\centerline{\bf Appendix B}
\medskip
\centerline{Recovering the Number Density and Scale Length
Distribution of LSBs \\
from the Observed Surface Density}
\bigskip

In this Appendix we investigate the relationship between the intrinsic
number density of objects at redshift $z$ with surface brightness
$\mu_0$ and physical scale length $R$, $N(\mu_0,R,z)$, and the observed
surface density of objects with apparent surface brightness $\mu$ and
angular scale length $\alpha$, $\Sigma(\mu,\alpha)$.  We also
calculate the underlying distribution of LSB scale lengths from
existing catalogs.

First, consider a volume element $dV(z)$ at an angular diameter
distance $D_a(z)$, subtending one steradian of the sky.  The number
density of objects in the shell that appear with angular scale length
$\alpha$ is

\begin{equation}
N(\mu_0,\alpha,z) = N(\mu_0,{R=\alpha D_a(z)},z) \, \frac{R}{\alpha},
\end{equation}

\ni where $\alpha$ is measured in radians.  If we allow objects to be
at large enough distances that surface brightness diminution plays a
role, then the observed surface brightness scales as $(1+z)^{-4}$ and
the number density becomes

\begin{equation}
N(\mu,\alpha,z) = N({\mu_0={\mu-10\log{(1+z)}}},\alpha,z) 
\end{equation}

\ni The number of objects per steradian per $\surfb$ is

\begin{equation}				\label{surfden}
\Sigma(\mu,\alpha) = \int^\infty_0 N(\mu,\alpha,z) dV(z).
\end{equation}

If a sample is close enough that all distances can be treated as
Euclidean, the volume element is $D_a^2dD_a=\alpha^{-3}R^2dR$, the
surface brightness of a galaxy does not change with distance, and
evolution in $N(\mu_0,\alpha,z)$ may be safely ignored.  Therefore,
for a uniform distribution of galaxies, Equation
\ref{surfden} becomes

\begin{equation}				\label{surfdenlocal}
\Sigma(\mu,\alpha) = \frac{1}{\alpha^4} \,
			\int^\infty_0 N({\mu_0=\mu},R)\,R^3\,dR.
\end{equation}

\ni Note that $\Sigma(\mu,\alpha)$ will always be proportional
to $\alpha^{-4}$, independent of the distribution of intrinsic scale
lengths.  Any size galaxy can be seen to a larger distance if it
is observed at a smaller limiting angular size; smaller values of $\alpha$
always probe larger volumes of the universe and always lead to
larger surface densities.

If we make the further assumption that $N(\mu_0,R)$ is separable, and
can be expressed as $N(\mu_0)f(R)$, then

\begin{equation}				\label{surfdenlocalsep}
\Sigma(\mu,\alpha) = \frac{N({\mu_0=\mu})}{\alpha^4} \,
			\int^\infty_0 f(R)\,R^3\,dR.
\end{equation}

\ni For an assumed form of $f(R)$, the fraction of objects with physical
scale length $R$, Equation \ref{surfdenlocalsep} may be inverted to
recover $N(\mu_0)$ from the observed distribution of surface densities
and angular scale lengths.  However, if $N(\mu_0,R)$ is not separable,
than the integral in equation \ref{surfdenlocalsep} is not independent
of $\mu_0$ and the relative proportion of galaxies in terms of
surface density $\Sigma(\mu,\alpha)$ does not reflect the relative
proportion of galaxies in real space.

The integral over $R$ diverges unless $f(R)$ falls off more quickly
than $R^{-3}$ for large $R$; this seems to be the case, as there are
no known galaxies to date which have scale lengths larger than
$100\kpc$.  We may uncover some confirmation for the cutoff at large
$R$ from the POSS-II survey for LSBs, a diameter-limited sample of
LSBs (Schombert et al.\ 1992) which includes several extremely large
galaxies similar to the giant LSB Malin-I (Bothun et al.\ 1987).  HI
observations of the ``V'' subsample of LSBs
($30\arcsec<\theta<1\arcmin$) have a 70\% detection rate, and imply a
$V/V_{max}$ statistic compatible with a uniform redshift distribution.
If $\Sigma({\cal R})d{\cal R}$ is the number of galaxies in a diameter
limited sample with physical radii at the isophotal diameter between
${\cal R}$ and ${\cal R}+d{\cal R}$, then the number density of
galaxies with physical radii between ${\cal R}$ and ${\cal R}+d{\cal
R}$ is

\begin{equation}				\label{POSSIInumden}
N({\cal R})\,d{\cal R} = 3\alpha_0^3 \, \frac{\Sigma({\cal R})}{{\cal R}^3}
\end{equation}

\ni where $\alpha_0\equiv[\alpha_{min}^{-3}-\alpha_{max}^{-3}]^{-1/3}$
and $\alpha_{min}$ and $\alpha_{max}$ are the diameter limits of the
survey.  The luminosity density is

\begin{equation}				\label{POSSIIlumden}
{\cal L}({\cal R})\,d{\cal R} = 3{\bar \Sigma}\alpha_0^3 \, \frac{\Sigma({\cal R})}{R},
\end{equation}

\ni assuming ${\bar \Sigma}$ is the mean surface brightness within
${\cal R}$.  These distributions are plotted in Figure
\ref{schombertVfig}.  Although the completeness and selection biases
of the Schombert et al.\ catalog are poorly understood, the strong
drop-off in luminosity density around ${\cal R}_{max}=10\kpc$ seems
quite robust; including the galaxies that were not detected in HI
could push the flat luminosity density out to ${\cal R}_{max}=20\kpc$,
but no further.  While there are some galaxies with rather large
sizes, the evidence suggests that they are neither an important
contributor to the number density or to the luminosity density.
McGaugh \& Bothun (1994) have done follow up observations of roughly a
dozen of the galaxies from the POSS-II survey which suggest that the
exponential scale length is typically one-quarter to one-half of the
isophotal radius.  This allows us to estimate the exponential scale
length $R$ given the isophotal radius ${\cal R}$.

The Schombert catalog data plotted in Figure \ref{schombertVfig}(b)
suggests that $f(R)$ is best described by a power law

\begin{equation}				\label{f_R}
f(R) = f_0 \, R^{-2}
\end{equation}

\ni between $R_{min}=0.2\kpc$ and $R_{max}=5\kpc$, where here $R$ is the
radius of the galaxy at the limiting isophote of the POSS-II survey.


\section{References}

\hi{Allen, R. J., \& Shu, F. H. 1979, \apj 227, 67.}

\hi{Bardeen, J. M., Bond, J. R., Kaiser, N., \& Szalay, A. S. 1986, \apj 304,
15.}

\hi{Bothun, G. D., Impey, C. D., \& Malin, D. F. 1991, \apj 376, 404.}

\hi{Bothun, G. D., Impey, C. D., Malin, D. F., \& Mould, J. 1987, \aj 94, 23.}

\hi{Bothun, G. D., Schombert, J. M., Impey, C. D., Sprayberry, D.,
McGaugh, S. S. 1993, \aj 106, 530.}

\hi{Caldwell, N., Armandroff, T. E., Seitzer, P., \& Da Costa, G. S.
1992, \aj 103, 840.}

\hi{Catelan, P. \& Theuns, T. 1996, to appear in \mn.}
 
\hi{Dalcanton, J. J. 1996, \apj 466, 92.}

\hi{Dalcanton, J. J., Spergel, D. N., \& Summers, F. J. 1997, \apj 482,
in press.}

\hi{Davies, J. I., Disney, M. J., Phillipps, S., Boyle, B. J., \& Couch, W.
J. 1994, \mn 269, 349.}

\hi{Davies, J. I., Phillipps, S., Cawson, M., Disney, M., \& Kibblewhite,
E. 1988, \mn 232, 239.}

\hi{de Blok, W. J. G., van der Hulst, J. M., \& Bothun, G. D. 1995, \mn 274,
235.}

\hi{de Jong, R. S. 1996, \aa, in press.}

\hi{de Jong, R. S., \& van der Kruit, P. C. 1994, \aasup 106, 451.}

\hi{Disney, M. J. 1976, \nat 263, 573.}

\hi{Disney, M., \& Phillipps, S. 1983, \mn 205, 1253.}

\hi{Driver, S. P. \& Phillipps, S. 1996, \apj 469, 529.}

\hi{Ellis, R. S., Colless, M., Broadhurst, T., Heyl, J., \& Glazebrook,
K. 1996, \mn 280, 235.}

\hi{Freeman, K. 1970, \apj 160, 811.}

\hi{Gunn, J. E., {\it et al.\ } 1987, {\it Opt.\ Eng.,} 26, 779.}

\hi{Impey, C., Bothun, G., \& Malin D. 1988, \apj 330, 634.}

\hi{Impey, C., Sprayberry, D., Irwin, M., \& Bothun, G. 1996, \apjs 105, 209.}

\hi{Irwin, M. J., Davies J. I., Disney, M. J., \& Phillips, S. 1990, \mn
245, 289.}

\hi{Jarvis, J. F., \& Tyson, J. A., 1981, \aj 86, 476.}

\hi{Knezek, P. 1993, Ph.D. Thesis, University of Massachusetts.}

\hi{Landolt, A. 1992, \aj 104, 340.}

\hi{Loveday, J., Peterson, B. A., Eftathiou, G., \& Maddox, S. J. 1992, \apj
390, 338.}

\hi{Marzke, R. O., Geller, M. J., Huchra, J. P., \& Corwin, H. G. 1994,
\aj 108, 437.}

\hi{McGaugh, S. S., \& Bothun, G. D. 1994, \aj 107, 530.}

\hi{McGaugh, S. S. 1994, \apj 426, 135.}

\hi{McGaugh, S. S. 1994, \nat 367, 538.}

\hi{McGaugh, S. S. 1995, in New Light on Galaxy Evolution, IAU Symp.\ 171,
ed.\ R.\ Bender \& R.\ Davies (Kluwer), in press.}

\hi{McGaugh, S. S. 1996, \mn 280, 337.}

\hi{Mo, H. J., McGaugh, S. S., \& Bothun, G. D. 1994, \mn 267, 129.}

\hi{Nilson, P. 1973, Uppsala Astr. Obs. Ann., 6, 1.}

\hi{Peterson, B. A., Ellis, R. S., Efstathiou, G., Shanks, T., Bean, A. K., Fong, R., \& Zen-Long, Z. 1986, \mn 221, 233}

\hi{Phillipps, S., Davies, J. I., \& Disney, M. J. 1990, \mn 242, 235.}

\hi{Postman, M., Lubin, L. M., Gunn, J. E., Oke, J. B., Schneider,
D. P., Hoessel, J. G., \& Christensen, J. A. 1996, \aj 111, 615.}

\hi{Roukema, B. F., \& Peterson, B. A. \aasup 109, 511.}

\hi{Sandage, A., \& Binggeli, B. 1984, \aj 89, 919.} 

\hi{Schneider, D. P., Schmidt, M., \& Gunn, J. E. 1994, \aj 107, 1245.}

\hi{Schombert, J. M., \& Bothun, G. D. 1988, \aj 95, 1389.}

\hi{Schombert, J. M., Bothun, G. D., Schneider, S. E., \& McGaugh, S. S. 1992,
\aj 103, 1107}

\hi{Schwartzenberg, J. M., Phillips, S., Smith, R. M, Couch, W. J., \&
Boyle, B. J. 1995, \mn 275, 121.}

\hi{Sprayberry, D., Impey, C., \& Irwin, M. J. 1996, \apj 463, 535.}

\hi{Sprayberry, D., Impey, C., Irwin, M. J. \& Bothun, G. D. 1997, \apj 481,
in press.}

\hi{Turner, J. A., Phillips, S., Davies, J. I., \& Disney, M. J. 1993, \mn
261, 39.}

\hi{Valdes, F. 1982, {\it S.P.I.E.}, 331, 465.}

\vfill
\clearpage


\begin{deluxetable}{llrrccclcl}

\scriptsize
\tablewidth{0pt}

\tablecaption{LSB Subsample\label{lsbtab}}

\tablehead{
\colhead{object}		&	\colhead{ID}		&
\colhead{RA} 			&	\colhead{Dec}		&
\colhead{$\mu_0(V)$}		&	\colhead{$\alpha$}	&
\colhead{$h$}			&	\colhead{$M_V$}		&
\colhead{$cz$}			&
\colhead{Spectral}		
\\
\colhead{}			&	\colhead{}		&
\colhead{(1950.0)} 		&	\colhead{(1950.0)}	&
\colhead{${\rm mag}/\Box\arcsec$}&	\colhead{($\arcsec$)}	&
\colhead{$h_{50}^{-1}\kpc$}	&	\colhead{}		&
\colhead{$\kms$}		&
\colhead{Features}		}

\startdata
PG 0849+4748 & R-27-1 &   8 49 28.4 & +47 48 37 & 23.1 & 4.3 & 3.6 & -18.3 &   8809 & H$\beta$,O[III],H$\alpha$\nl
PG 0847+4747 & R-26-1 &   8 47 44.9 & +47 47 48 & 23.2 & 2.6 &  -  &    -  &
 -   & \nl
PG 1521+4632 & M-232-1 & 15 21 32.4 & +46 32 58 & 23.6 & 3.2 & 1.7 & -16.1 &   5600: & break?\nl
PG 1136+4750 & Q-129-2 & 11 36 46.1 & +47 50 41 & 23.7 & 4.5 & 1.9 & -16.2 &   4270 & H$\beta$,O[III],H$\alpha$\nl
PG 0914+4744 & Q-42-1  &  9 14 24.5 & +47 44 48 & 23.9 & 6.1 & 3.0 & -17.1 &   5400 & break,H$\beta$?\nl
PG 1133+4755 & R-127-1 & 11 33 35.4 & +47 55 26 & 24.7 & 6.9 & 3.6 & -16.7 &   5250 & break\nl
PG 1327+4637 & M-161-1 & 13 27 46.2 & +46 37 47 & 24.8 & 6.2 & 3.5 & -16.5 &   5600 & break\nl

\enddata

\end{deluxetable}

\vfill
\clearpage


\figcaption[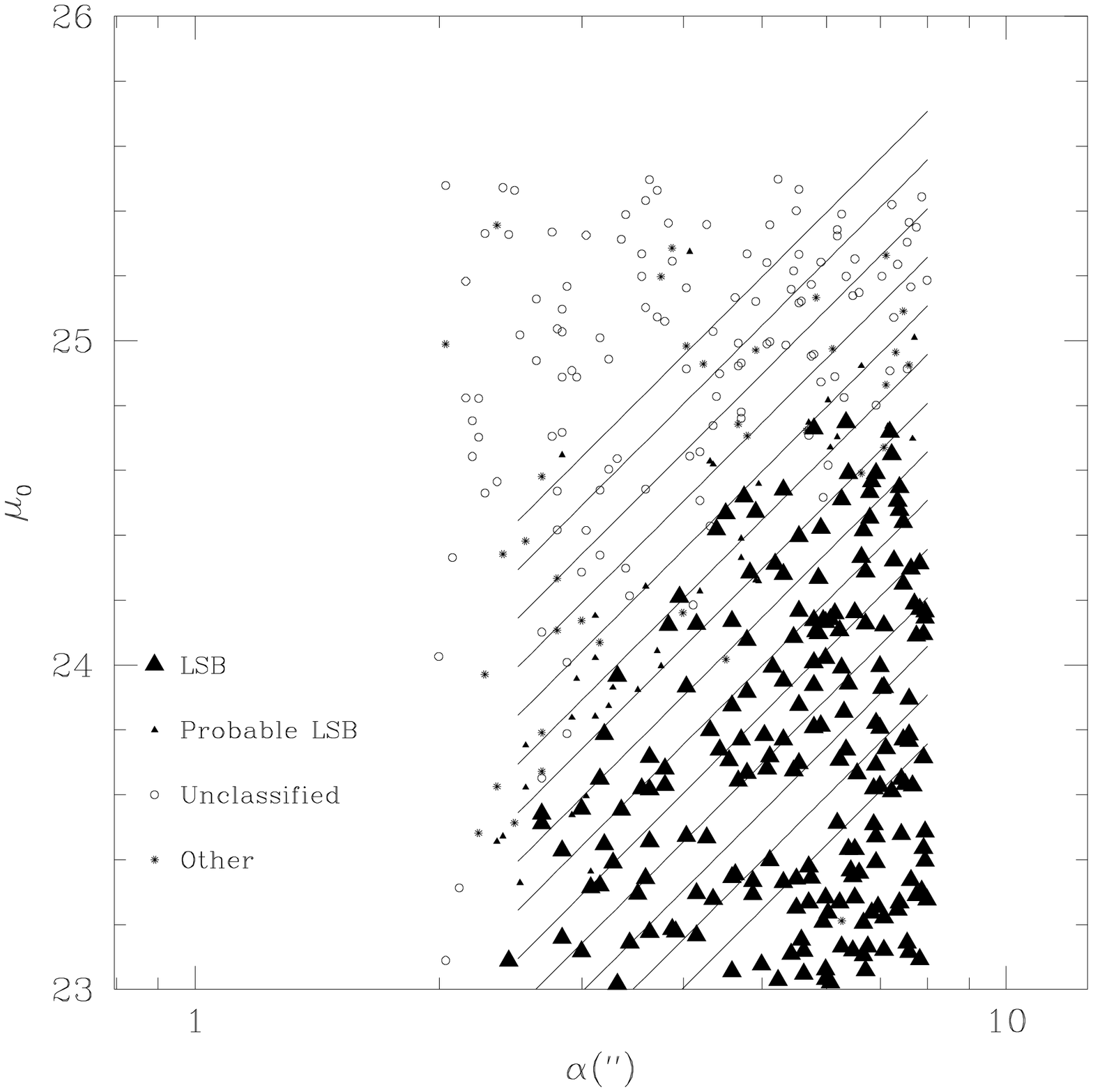,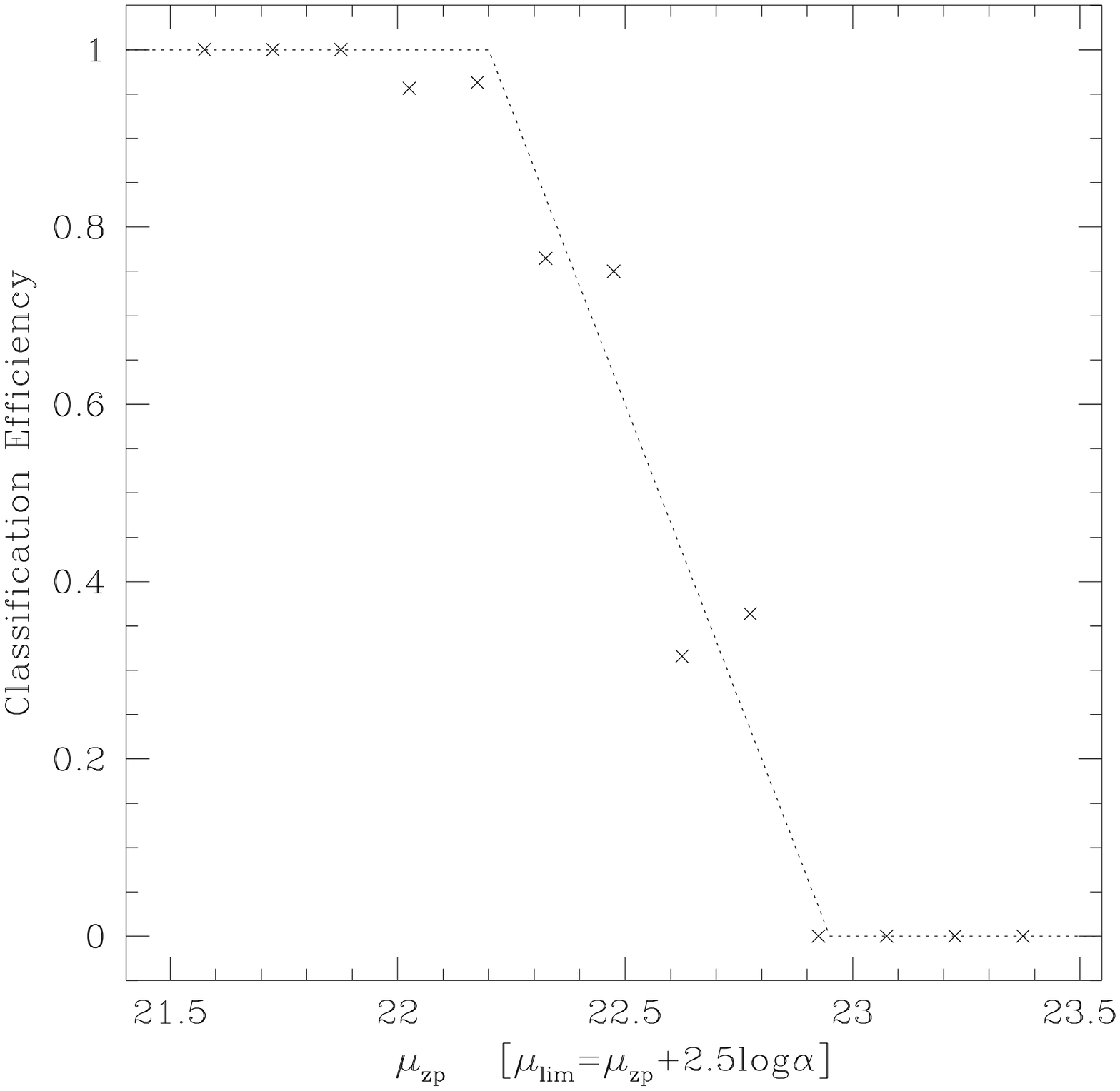] {
(a) Classification of artificial LSBs as a function of exponential
disk profile $\alpha$ and central surface brightness $\mu_0$.  The
symbols represent the classification assigned to each artificial
galaxy: LSB ({\it large triangle}), possible LSB ({\it small triangle}),
other classification(s) ({\it asterix}), no classification ({\it open circle}).
The solid lines are lines of constant signal-to-noise (i.e. varying
values of $\mu_{zp}$).\\
(b) Classification efficiency vs $\mu_{zp}$.  The fraction of LSBs
from Figure \ref{classfig}[a] which were correctly identified as
LSBs, as a function of $\mu_{zp}$ (solid lines in Figure \ref{classfig}[a]).
The dotted line is the fit given in equation \ref{classeqn}.
\label{classfig}}

\figcaption[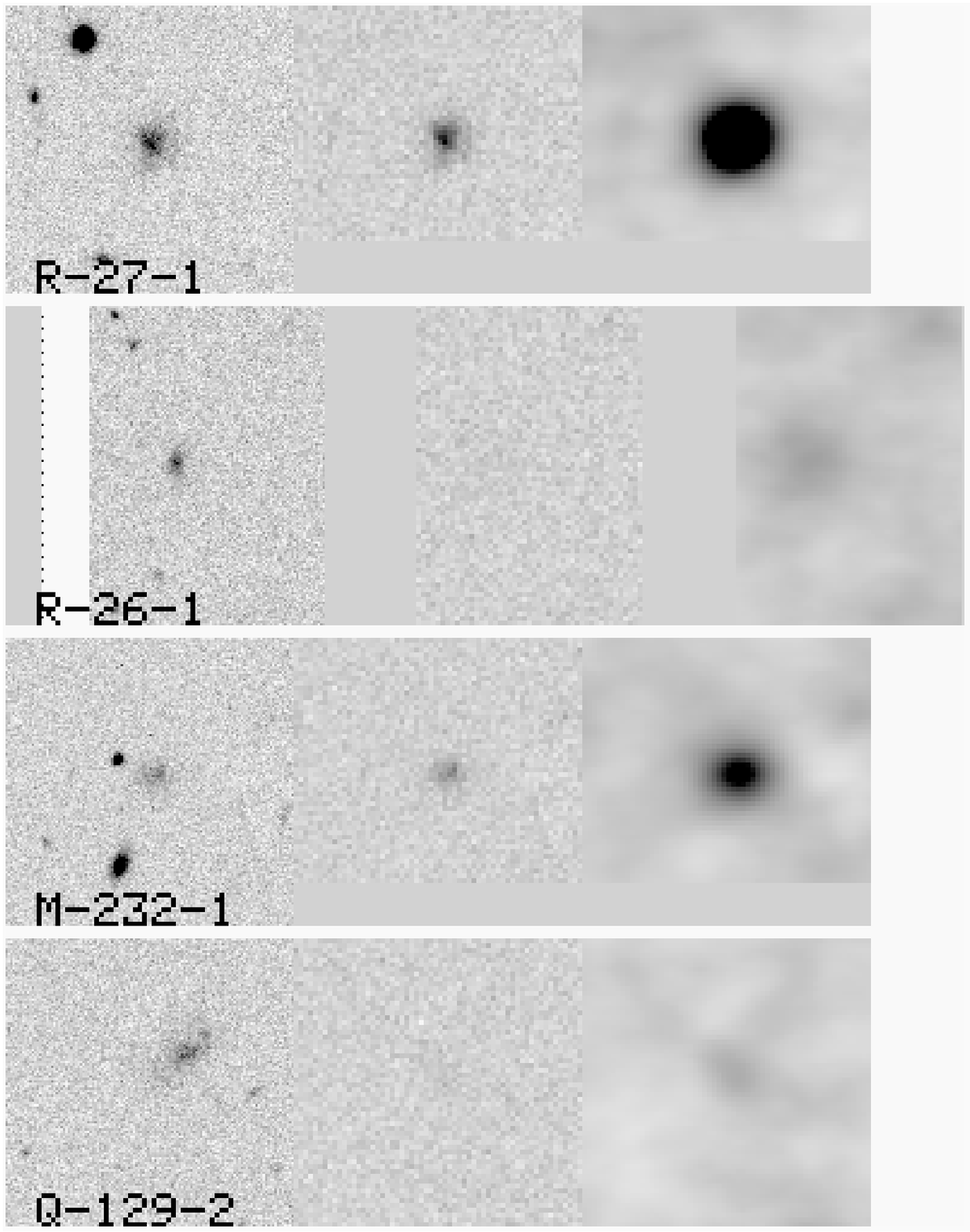,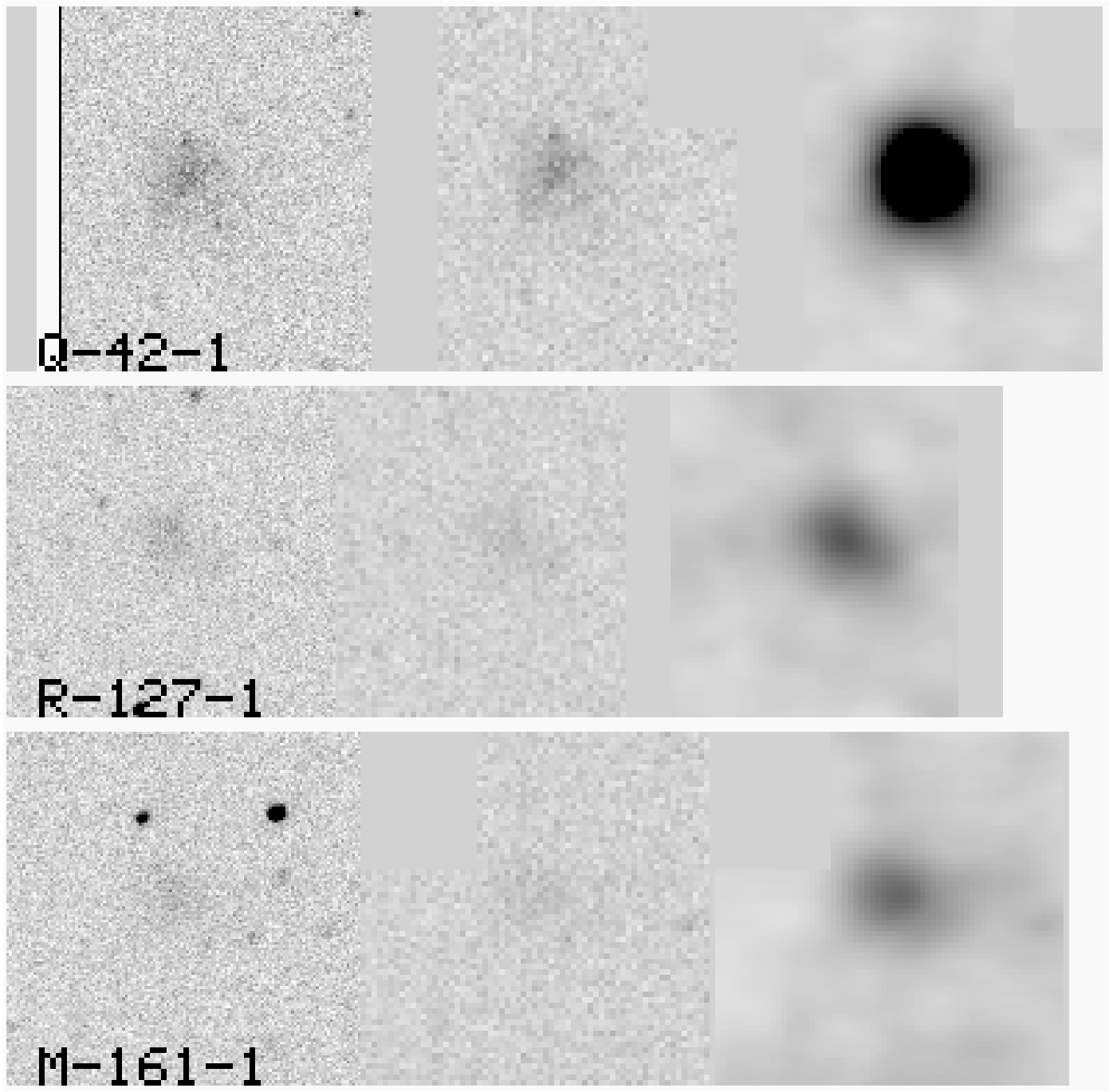] {
Images of the LSBs in the final subsample.  Each image constists
of three panels taken from different stages of
processing, centered on the position of the candidate and adjusted to
a common angular size.  The first panel is taken from the data before
cleaning.  The second panel is taken from the data after cleaning,
flattening, and rebinning by a factor of two.  The third panel is the
smoothed image used for detection.  Because of cleaning, high surface
brightness features detectable in the leftmost panel may not
be visible in the central cleaned image, although there is sufficient
diffuse emission which is uncleaned for the galaxy to be detected
in the smoothed image.
\label{lsbfig}}

\figcaption[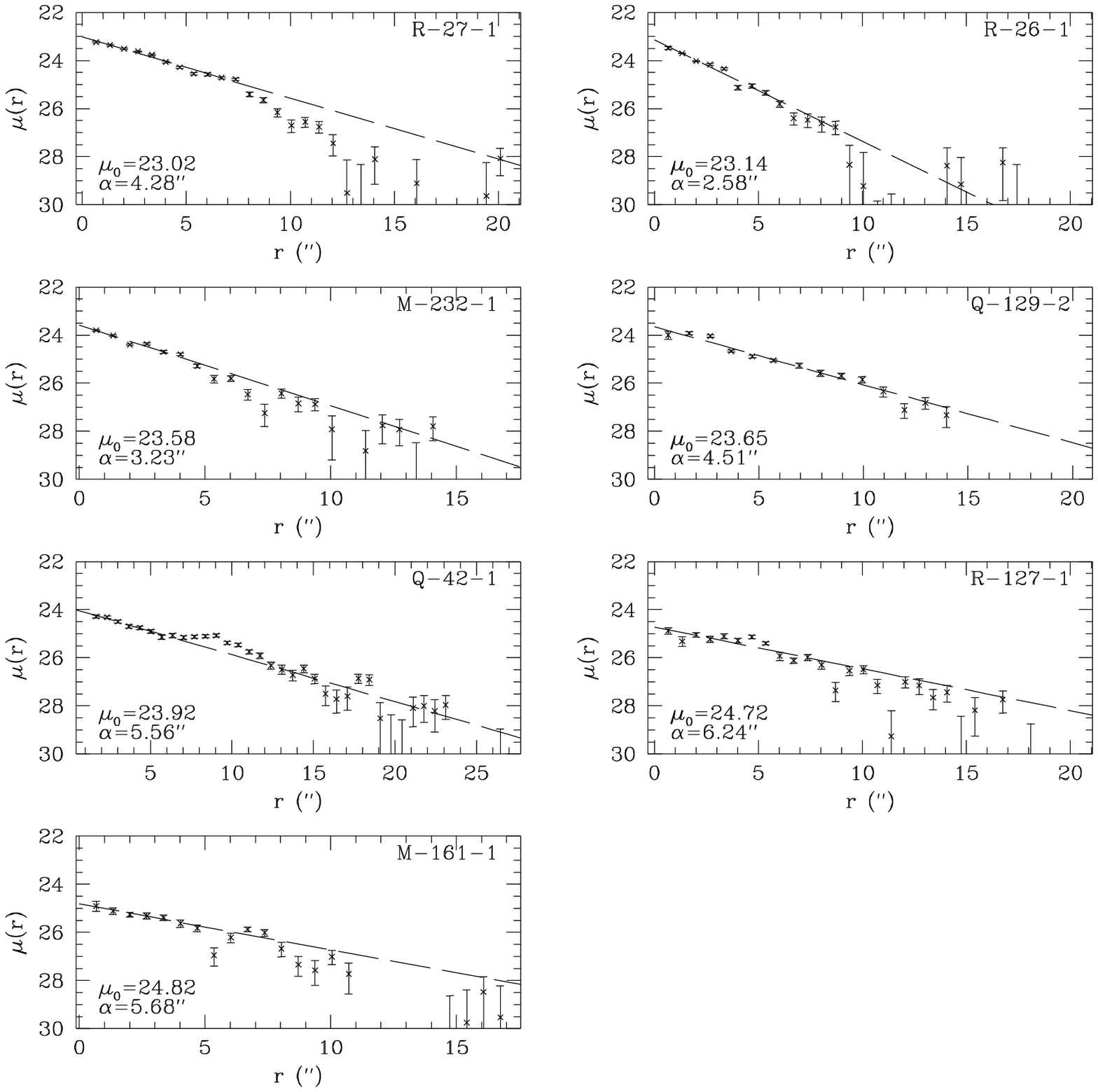] {
Radial surface brightness profiles for the LSBs in the final
subsample.  Dashed lines are fits of exponential surface
brightness profiles to the measured intensity.  The error bars
include the photometric uncertainty at each radius, as well as the
uncertainty in the sky subtraction.
\label{profilefig}}

\figcaption[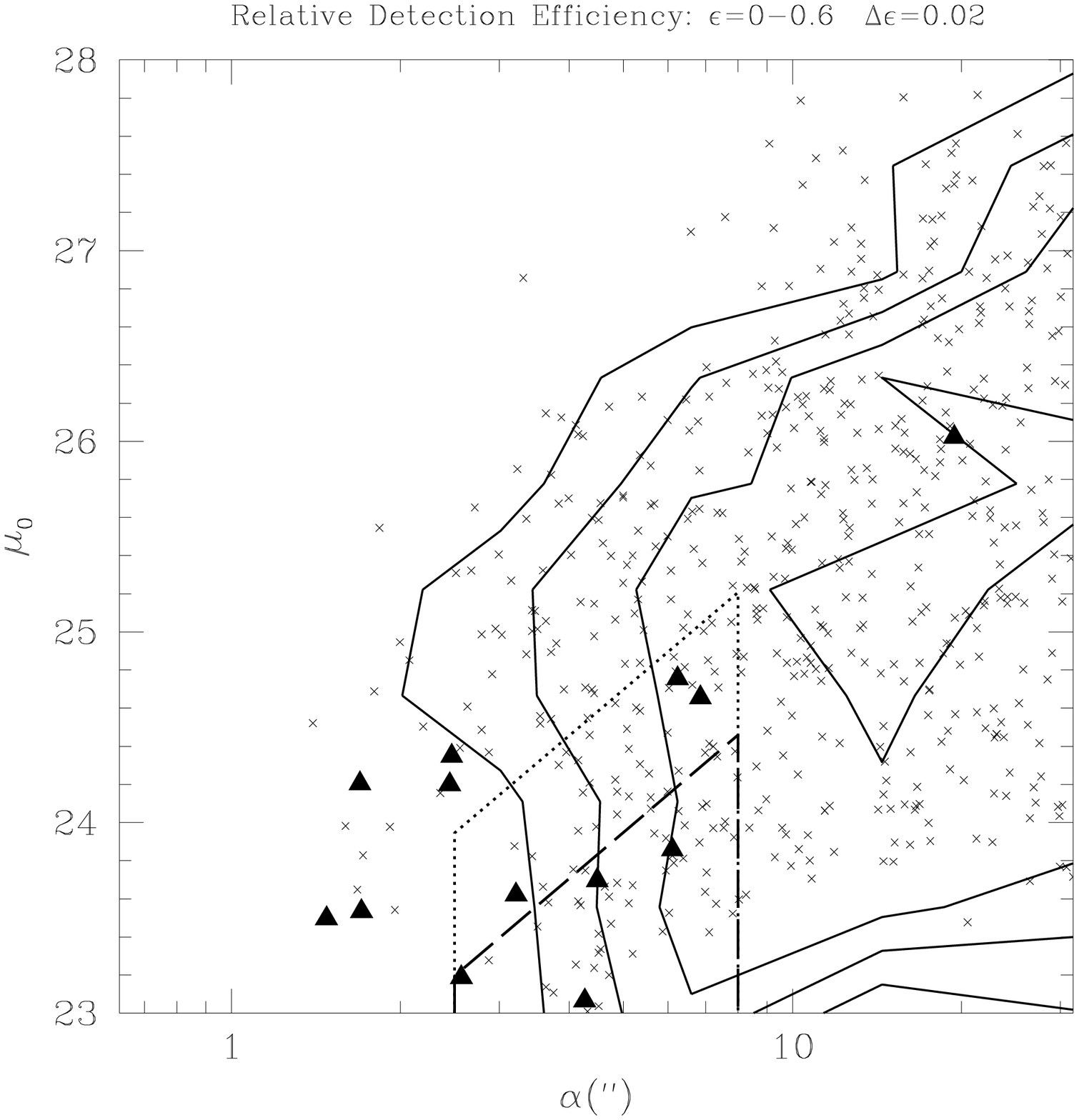,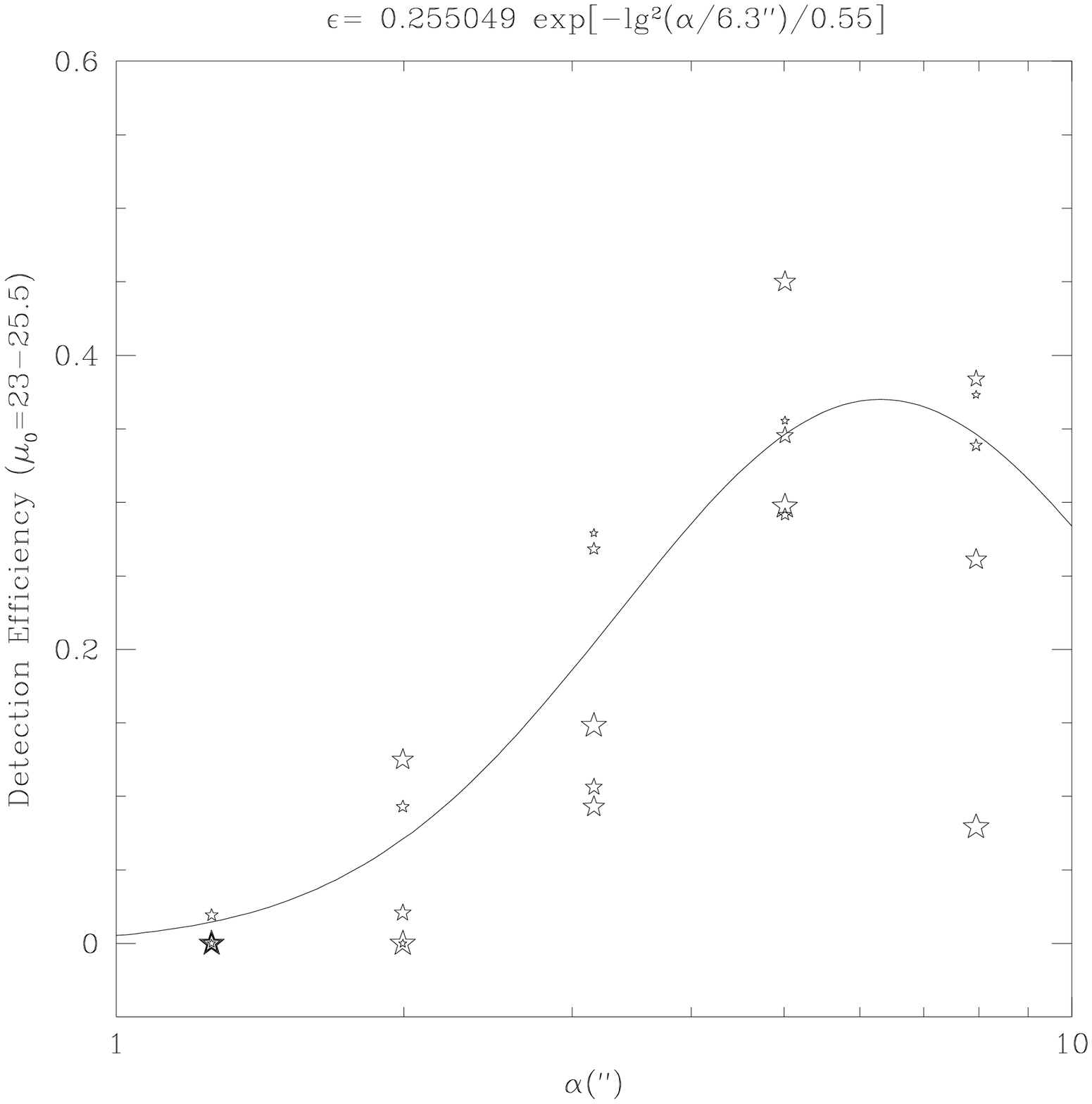] {
(a) Detection efficiency of artificial LSBs as a function of
exponential disk profile $\alpha$ and central surface brightness
$\mu_0$.  The contours are lines of constant detection efficiency,
separated by 0.1.  The locations of the artificial galaxies which were
successfully detected in Monte Carlo simulations are plotted as small
$\times$'s.  The triangles are LSB candidates with $\mu_0>23\surfb$ in
$V$.  The dashed line is the region of 100\% classification
efficiency, and the dotted line is where the classification efficiency
approaches zero.\\
(b) Binned detection efficiency as a function of exponential
disk profile $\alpha$, for $\mu_0$ between 23 and $25.5\surfb$ in $V$,
in steps of 0.5; the symbol size increases with increasing surface
brightness.  The solid line is the gaussian approximation given in
equation \ref{detectioneqn}.
\label{detectionfig}}

\figcaption[figure5.ps,figure5_cont.ps] {
Spectra of individual LSBs (dark lines, lower spectrum), compared with
the template LSB spectra which were used for cross-correlation (light
lines, upper spectrum) and which were generated by averaging the
spectra from all other LSBs, excluding R-26-1.  Spectra have been
shifted to their rest-wavelengths; we have shifted the uncertain
R-26-1 spectrum by $cv=17200\kms$.  For comparison, the final panel
compares one composite LSB spectra with a K1III giant observed with
the same instrumental arrangement after a wavelength dependent flux
correction has been applied.  All spectra have been smoothed with a 3
pixel boxcar filter, except R-26-1, M-232-1, and M-161-1, which were
smoothed with a 5 pixel boxcar filter.  The coadded LSB spectra have
been given an arbitrary vertical offset to facilitate comparison.  The
[OII] line is broadened, due to an inability to maintain the focus the
spectrograph in the UV.  Obvious sky subtraction errors at bright sky
lines have been interpolated over, for clarity of presentation.
\label{spectrafig}}

\figcaption[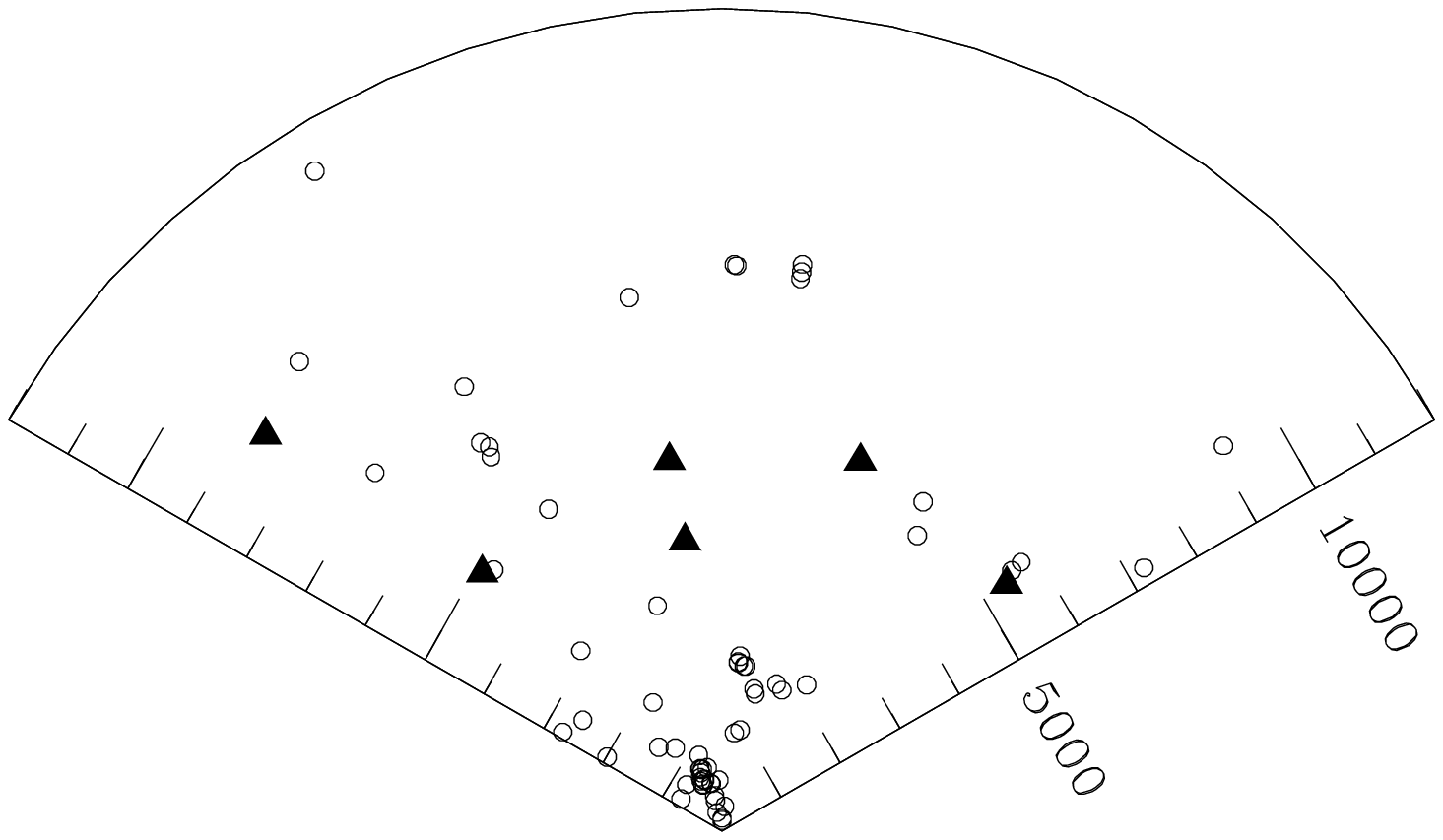] {
Space-Redshift diagram for the LSBs in the final subsample ({\it
triangles}, excluding R-26-1) and normal galaxies ({\it circles})
taken from the ZCAT redshift catalog.  The normal galaxies are taken
from $1\deg$ wide strips in declination, spanning the range of right
ascensions mapped by the survey.  At $5500\kms$, the width of
the $1\deg$ strip is $1.9\,h_{50}^{-1}\Mpc$.
\label{piefig}}

\figcaption[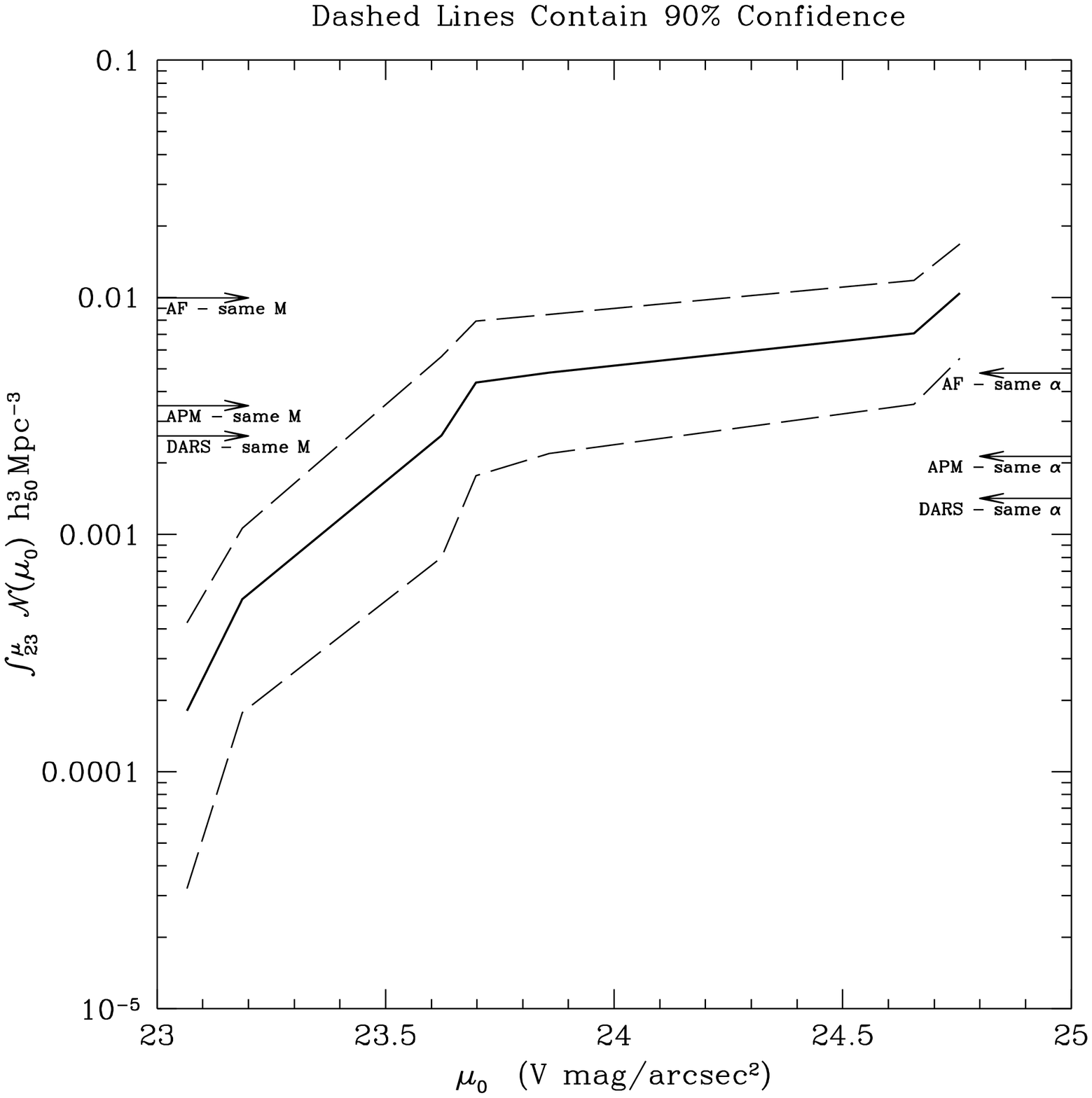] {
The integrated number density of LSB galaxies with $\mu_0>23\surfb$ in
$V$.  The dashed lines encompass the 90\% confidence interval.  The
arrows are drawn at the integrated number density of normal
galaxies, as cataloged by the DARS survey (Peterson et al.\ 1986), the
Stromlo-APM survey (APM; Loveday et al.\ 1992), and the AutoFib survey
(AF; Ellis et al.\ 1996).  The range of integration was restricted to
either the same scale lengths (``same $\alpha$''; right axis) or
same absolute magnitudes (``same M''; left axis) as the LSBs in
the survey.  
\label{numdenfig}}

\figcaption[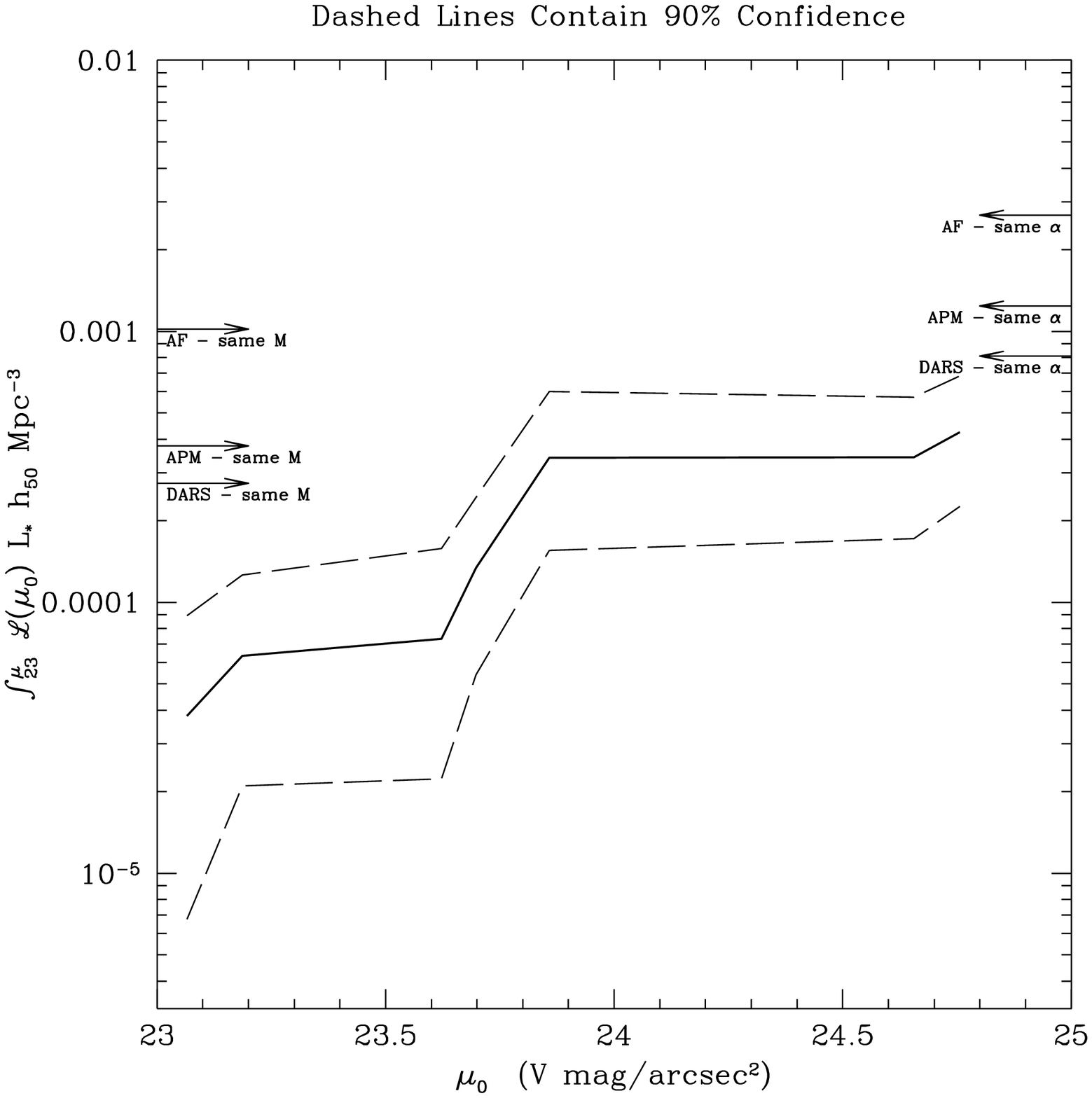] {
Similar to Figure \ref{numdenfig}, except the integrated
luminosity density of LSB galaxies with $\mu_0>23\surfb$ in $V$ is
plotted, in units of $L_* h_{50} Mpc^{-3}$, chosen to correspond to
$M_{b_j}=-19.5$.  The dashed lines encompass the 90\% confidence
interval.  The arrows are drawn at the luminosity density of normal
galaxies, as cataloged by the DARS survey (Peterson et al.\ 1986), the
Stromlo-APM survey (APM; Loveday et al.\ 1992), and the AutoFib survey
(AF; Ellis et al.\ 1996).  The range of integration was restricted to
either the same scale lengths (``same $\alpha$''; right axis) or same
absolute magnitudes (``same M''; left axis) as the LSBs in the survey.
\label{lumdenfig}}

\figcaption[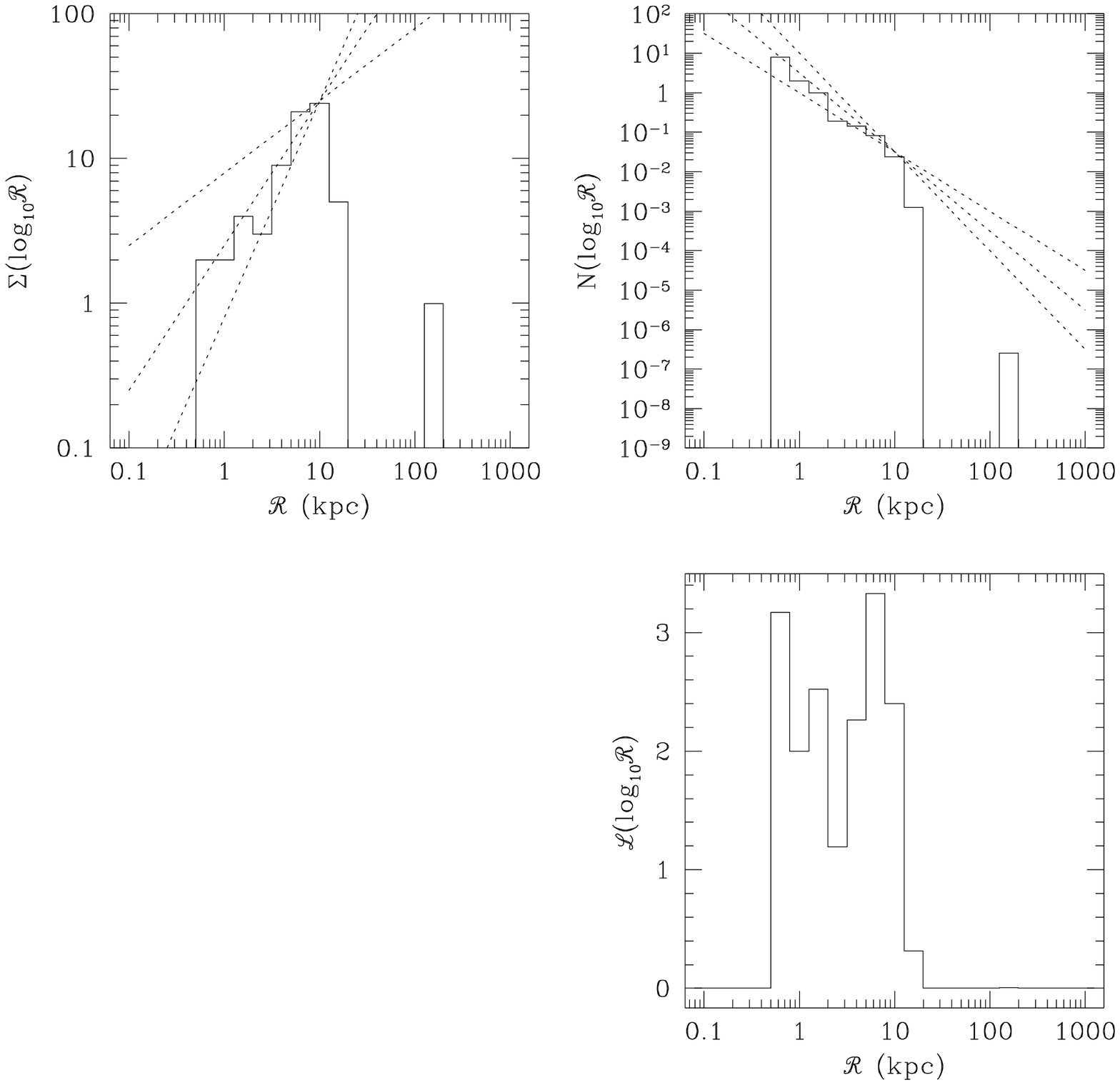] {
The distribution of radii for the Schombert et al.\ (1992)
``V'' catalog of LSBs, and the implied relative number and luminosity
density of LSBs as a function of physical size.  The angular diameters
and distances for the LSBs have been converted to physical radii,
making no correction for the dependence of angular size on surface
brightness.  (a) The number of objects per $0.2\log{(R/\kpc)}$ in the
catalog.  There are 71 LSBs in the distribution. There are an
additional 29 galaxies which were observed in HI but not detected.
The dotted lines are drawn for power-law slopes of 0.5, 1, and 1.5.
The distribution is consistent with slopes of 1 and 1.5, but not 0.5.
(b) The relative number density of galaxies per $0.2\log{(R/\kpc)}$,
calculated by correcting the distribution in (a) for the size of the
volume in which the LSBs could have been identified, given the
diameter limits of the survey and the physical sizes of the galaxies
at their diameter limits (see eqn. \ref{POSSIInumden}).  The dotted
lines are drawn for power-law slopes of -1.5, -2, and -2.5.  (c) The
relative luminosity density of LSBs per $0.2\log{(R/\kpc)}$,
calculated from (b) assuming a constant mean surface brightness within
$R$, so that luminosity is proportional to $R^2$ (see
eqn. \ref{POSSIInumden}).  The luminosity density is relatively flat
between $0.5\kpc$ and $10\kpc$, but drops precipitously at larger
sizes.  The LSBs in the ``V'' catalog do not have well determined
selection criteria; as such, these plots should not be taken as
definitive measurements, but as rough estimates.  They have central
surface brightnesses $<24\,B\surfb$, brighter than almost all of the
candidates in our survey.
\label{schombertVfig}}

\vfill
\clearpage

{	

\pagestyle{empty}

\begin{figure}[p]
\centerline{ \psfig{figure=figure1a.ps} }
\begin{flushright}{\bigskip\cap Figure 1[a]}\end{flushright}
\end{figure}
\vfill
\clearpage
\begin{figure}[p]
\centerline{ \psfig{figure=figure1b.ps} }
\begin{flushright}{\bigskip\cap Figure 1[b]}\end{flushright}
\end{figure}
\vfill
\clearpage

\begin{figure}[p]
\centerline{ \psfig{figure=figure2.ps} }
\begin{flushright}{\bigskip\cap Figure 2}\end{flushright}
\end{figure}
\vfill
\clearpage
\begin{figure}[p]
\centerline{ \psfig{figure=figure2_cont.ps} }
\begin{flushright}{\bigskip\cap Figure 2 (continued)}\end{flushright}
\end{figure}
\vfill
\clearpage

\begin{figure}[p]
\centerline{ \psfig{figure=figure3.ps} }
\begin{flushright}{\bigskip\cap Figure 3}\end{flushright}
\end{figure}
\vfill
\clearpage

\begin{figure}[p]
\centerline{ \psfig{figure=figure4a.ps} }
\begin{flushright}{\bigskip\cap Figure 4[a]}\end{flushright}
\end{figure}
\vfill
\clearpage
\begin{figure}[p]
\centerline{ \psfig{figure=figure4b.ps} }
\begin{flushright}{\bigskip\cap Figure 4[b]}\end{flushright}
\end{figure}
\vfill
\clearpage


\begin{figure}[t]
{ \begin{minipage}[l]{3.85in}
	\psfig{file=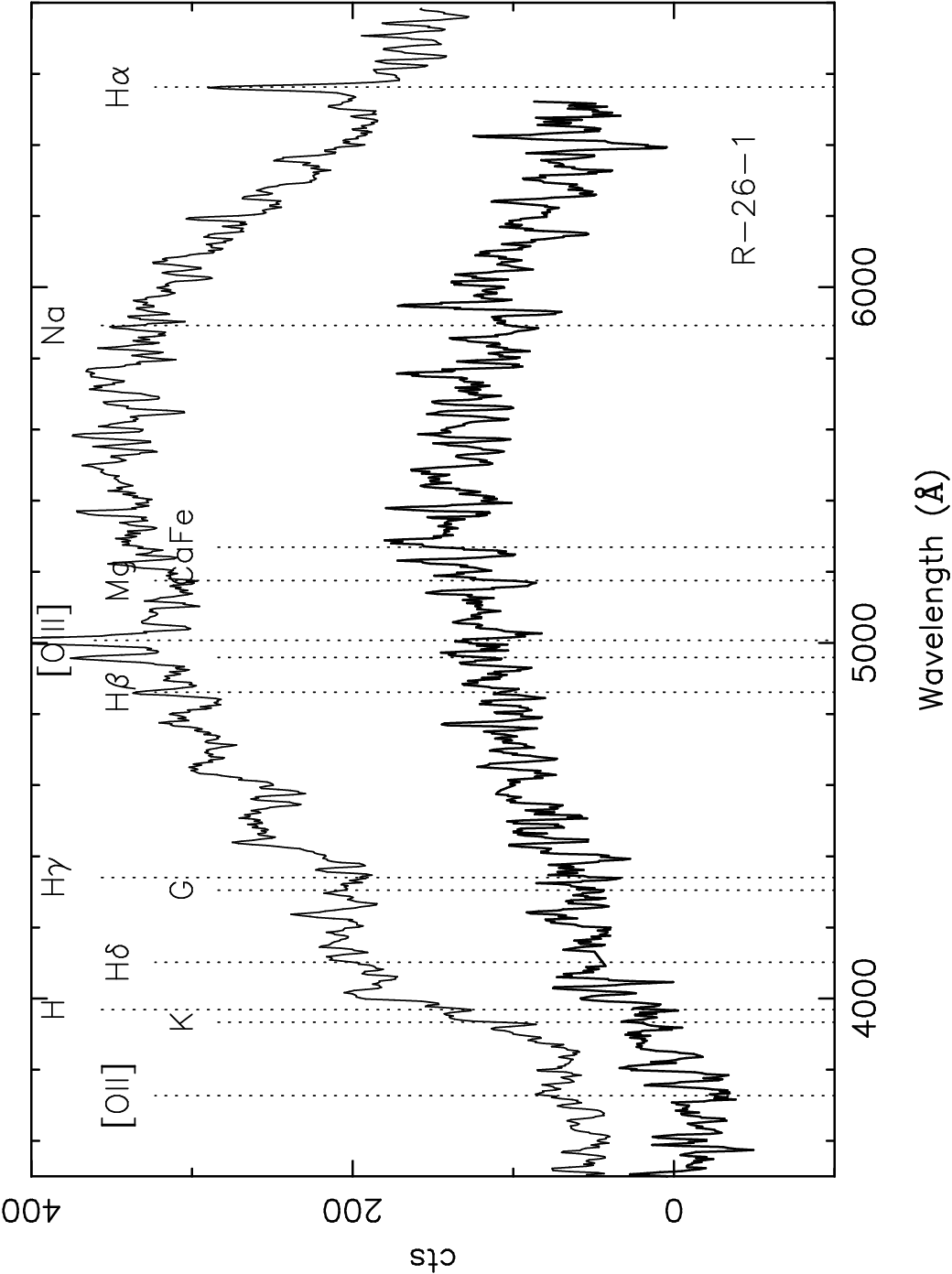,height=3.85in,width=3.19in}
  \end{minipage}   \ \hfill \
  \begin{minipage}[r]{3.85in}
	\psfig{file=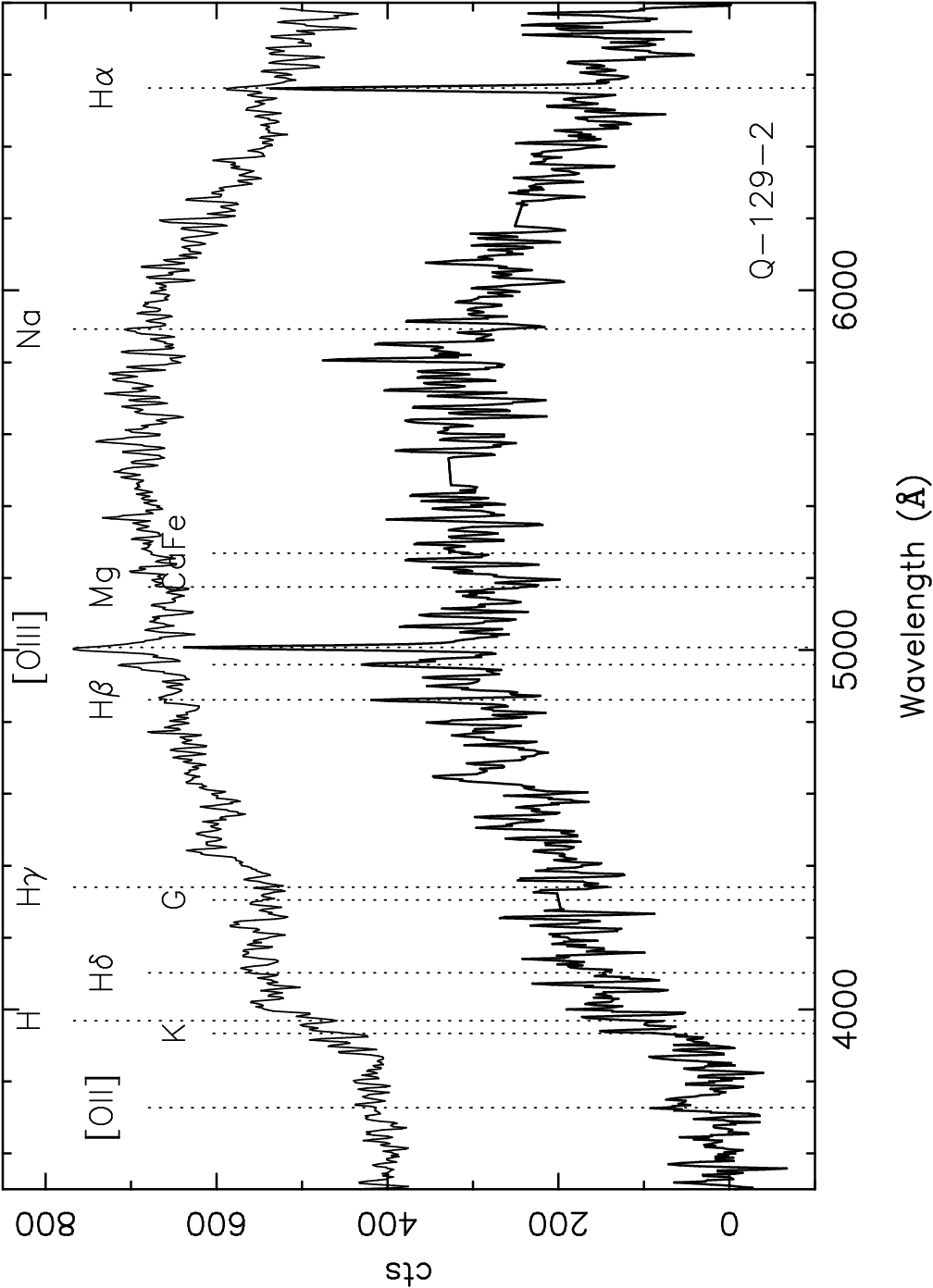,height=3.85in,width=3.19in}
  \end{minipage}}
\end{figure}
\begin{figure}[b]
{ \begin{minipage}[l]{3.85in}
	\psfig{file=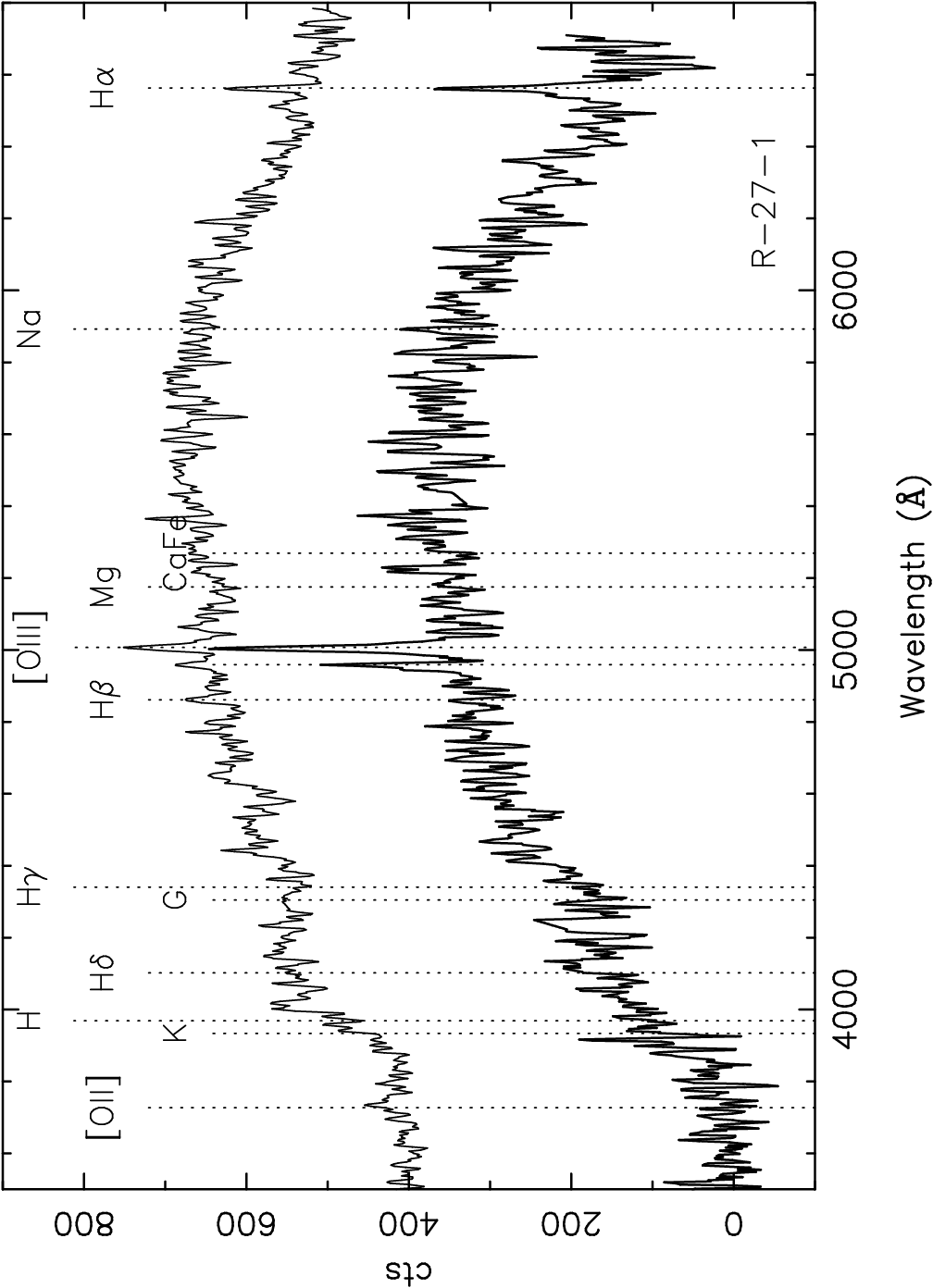,height=3.85in,width=3.19in}
  \end{minipage}   \ \hfill \
  \begin{minipage}[r]{3.85in}
	\psfig{file=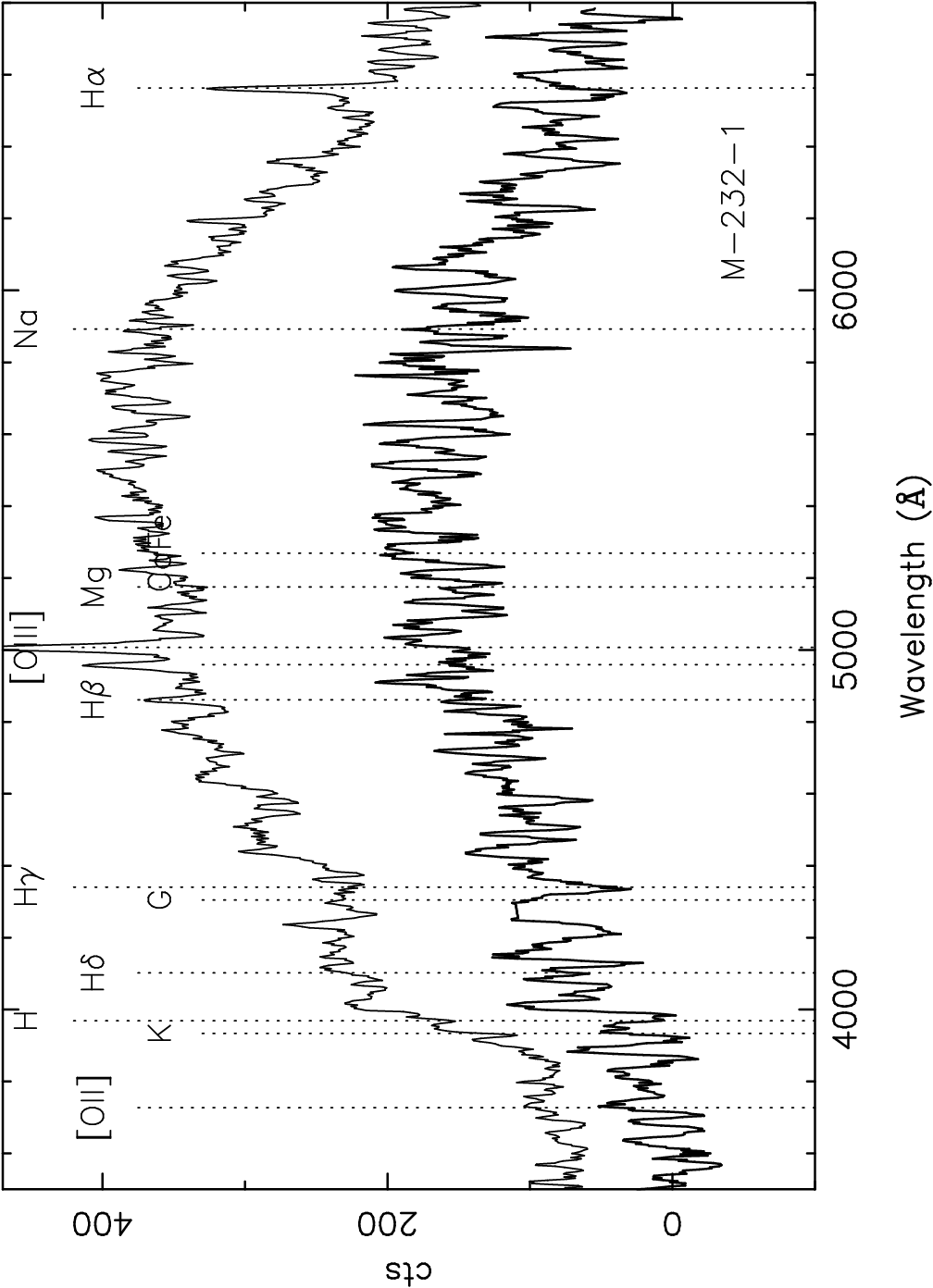,height=3.85in,width=3.19in}
  \end{minipage}}
\end{figure}
\clearpage

\begin{figure}[t]
{ \begin{minipage}[l]{3.75in}
	\psfig{file=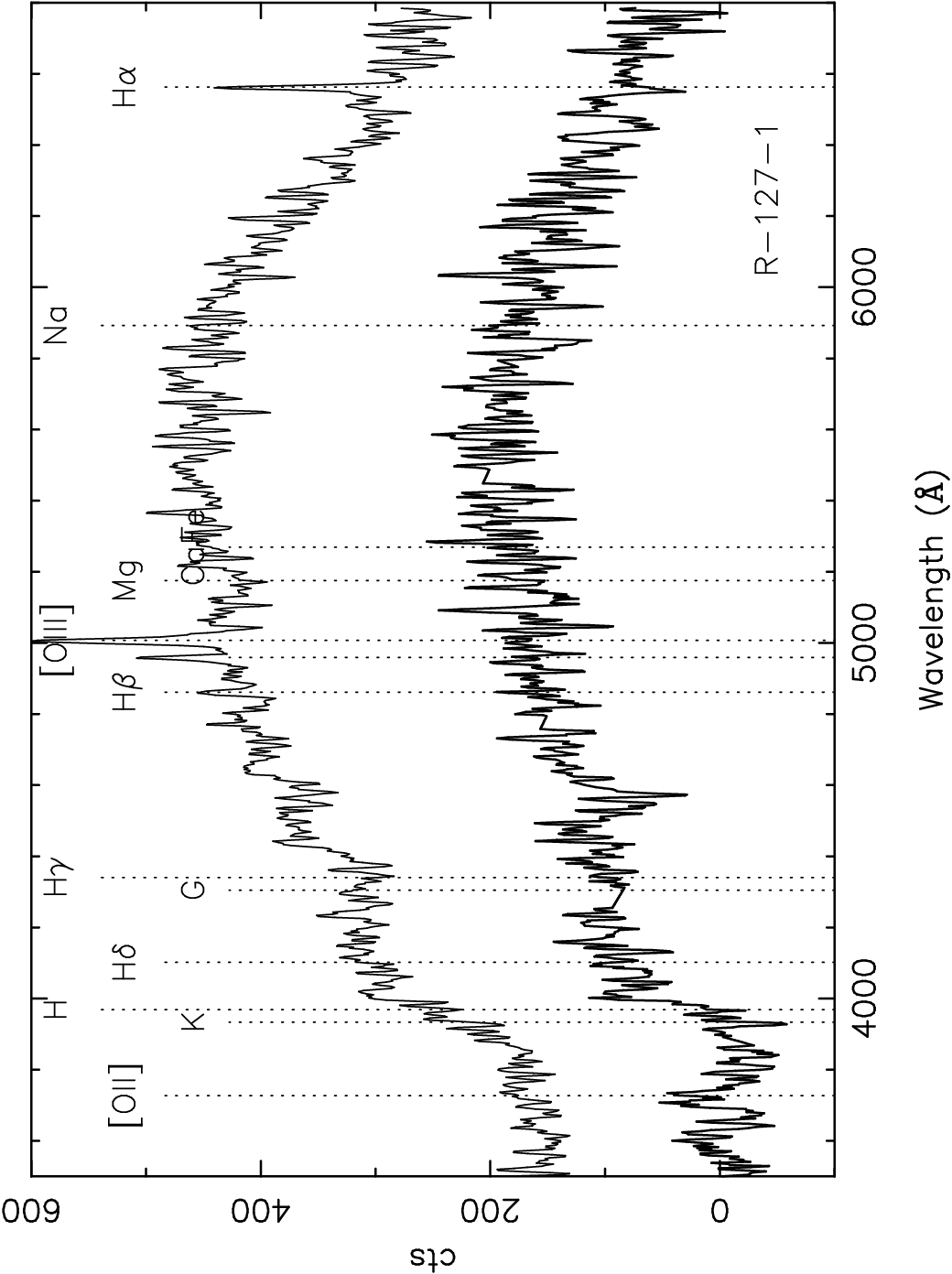,height=3.80in,width=3.19in}
  \end{minipage}   \ \hfill \
  \begin{minipage}[r]{3.75in}
	\psfig{file=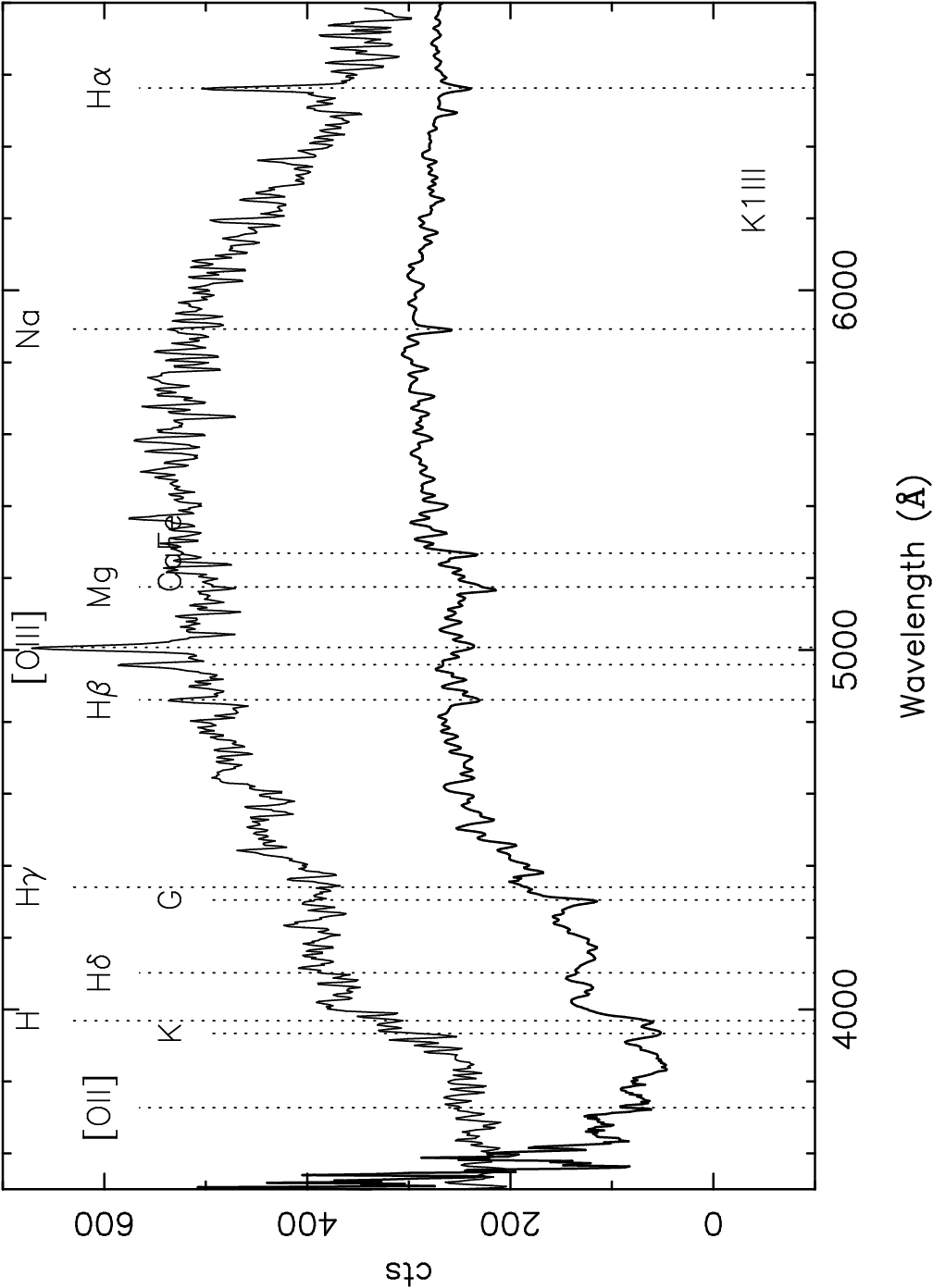,height=3.80in,width=3.19in}
  \end{minipage}}
\end{figure}
\begin{figure}[b]
{ \begin{minipage}[l]{3.75in}
	\psfig{file=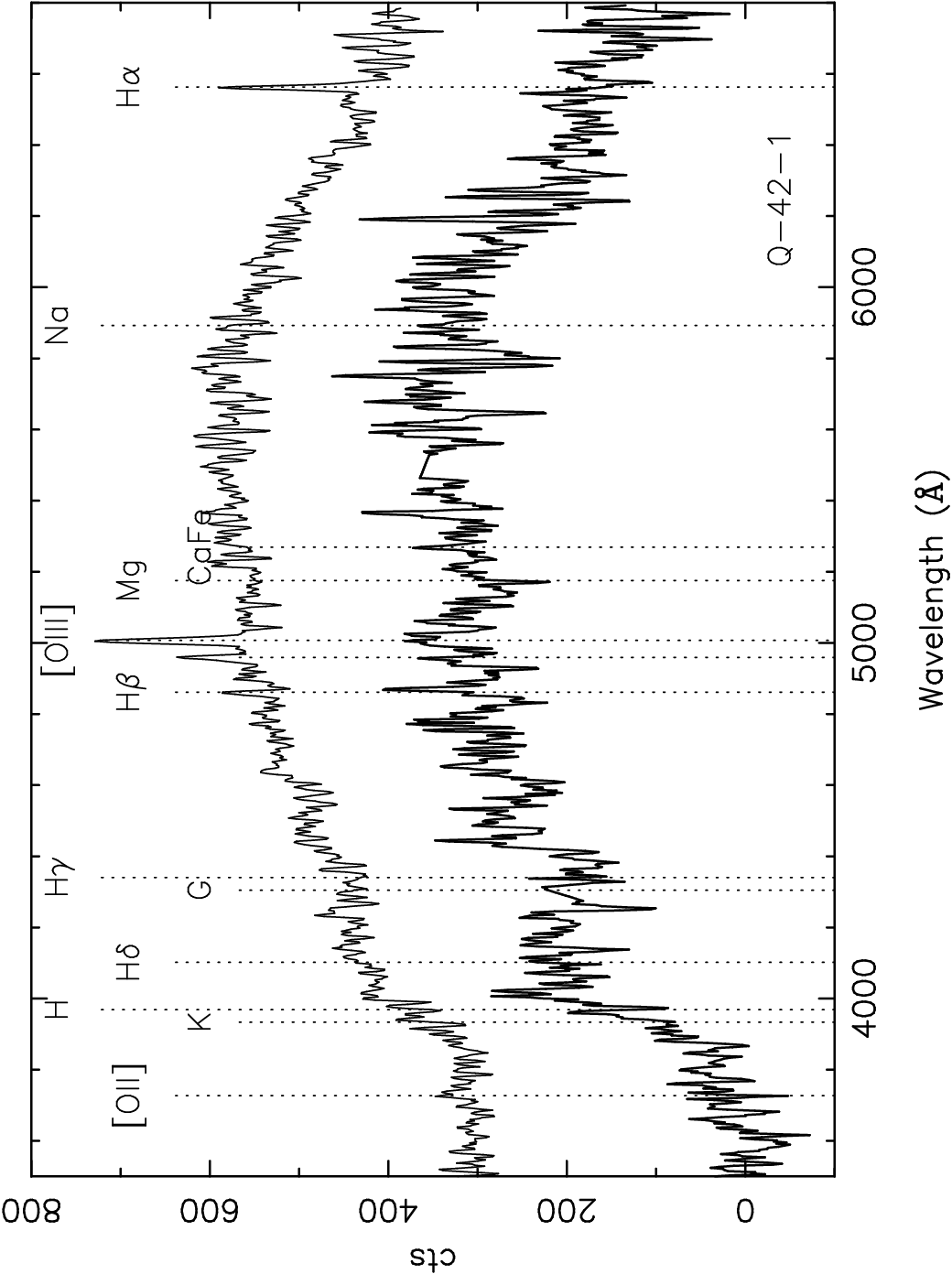,height=3.80in,width=3.19in}
  \end{minipage}   \ \hfill \
  \begin{minipage}[r]{3.75in}
	\psfig{file=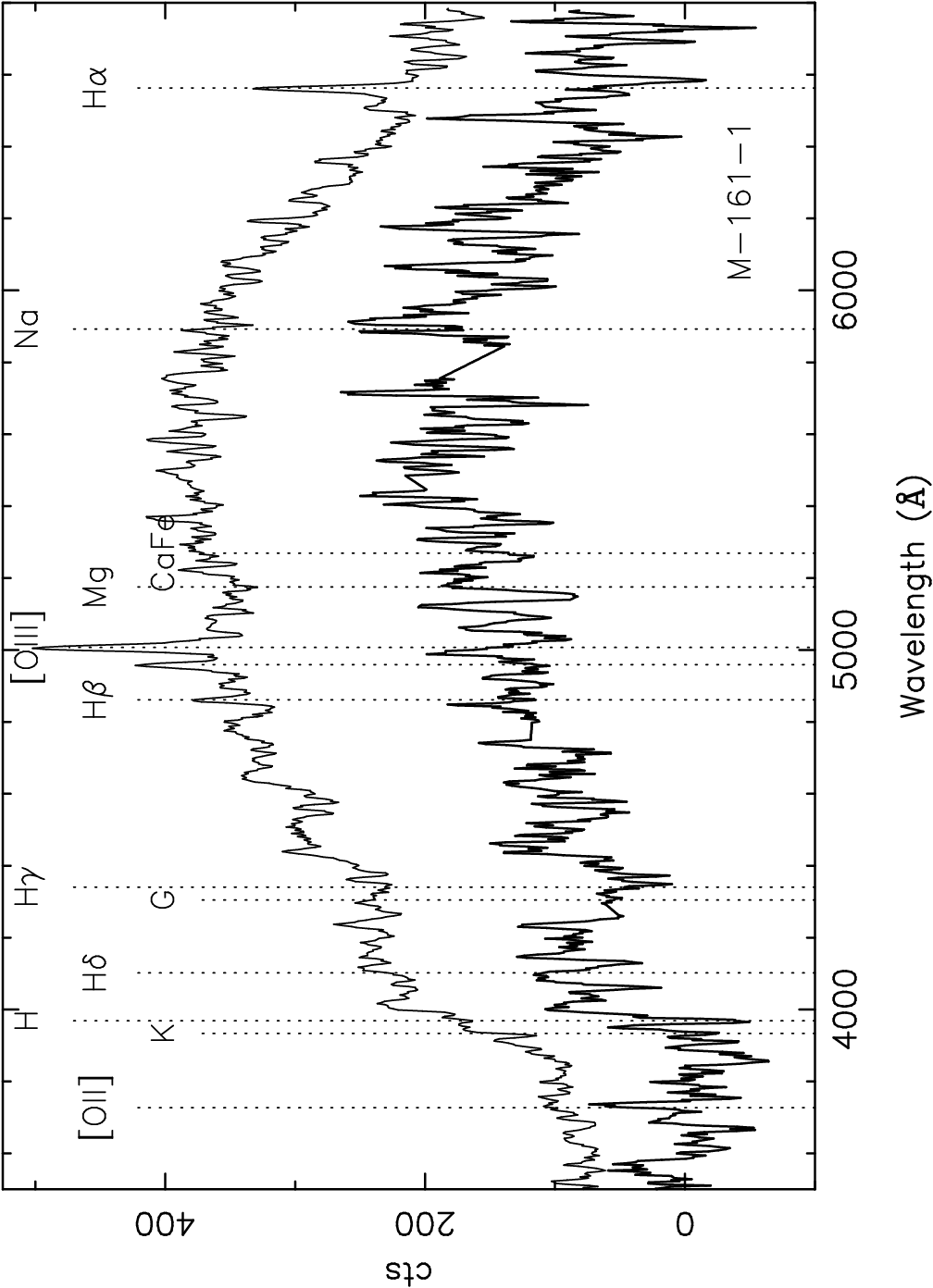,height=3.80in,width=3.19in}
  \end{minipage}}
\end{figure}
\clearpage


\begin{figure}[p]
\centerline{ \psfig{figure=figure6.ps} }
\begin{flushright}{\bigskip\cap Figure 6}\end{flushright}
\end{figure}
\vfill
\clearpage

\begin{figure}[p]
\centerline{ \psfig{figure=figure7.ps} }
\begin{flushright}{\bigskip\cap Figure 7}\end{flushright}
\end{figure}
\vfill
\clearpage

\begin{figure}[p]
\centerline{ \psfig{file=figure8.ps} }
\begin{flushright}{\bigskip\cap Figure 8}\end{flushright}
\end{figure}
\vfill
\clearpage

\begin{figure}[p]
\centerline{ \psfig{file=figure9.ps} }
\begin{flushright}{\bigskip\cap Figure 9}\end{flushright}
\end{figure}
\vfill
\clearpage
}
\end{document}